\documentclass[a4paper,11pt]{article}
\pdfoutput=1 
\usepackage{jcappub}
\usepackage{graphicx} % Required for inserting images
\usepackage{empheq}

\usepackage[utf8]{inputenc}
\pdfoutput=1 

\usepackage{graphicx}
\usepackage[figuresright]{rotating}
\usepackage{bm}
\usepackage{color}
\usepackage{placeins}
\usepackage{textcomp}
\usepackage{multirow}
\usepackage{hyperref}
\usepackage{ulem}
\usepackage{accents}
\usepackage{xspace}

\usepackage{changes}
\usepackage{amssymb}
\usepackage{amsmath}

\definecolor{dgreen}{rgb}{0,0.6,0.0}

\newcommand{\HH}{{\cal H}}

\newcommand{\al}{\alpha}

\newcommand{\de}{\delta}
\newcommand{\De}{\Delta}
\newcommand{\ep}{\epsilon}

\newcommand{\ga}{\gamma}

\newcommand{\ka}{\kappa}

\newcommand{\La}{\Lambda}
\newcommand{\Om}{\Omega}
\newcommand{\om}{\omega}
\newcommand{\si}{\sigma}

\newcommand{\vth}{\vartheta}

\newcommand{\ra}{\rightarrow}

\newcommand{\bnabla}{{\boldsymbol\nabla}}
\newcommand{\be}{\begin{equation}}
\newcommand{\ee}{\end{equation}}

\newcommand{\bea}{\begin{eqnarray}}
\newcommand{\eea}{\end{eqnarray}}
\newcommand{\bean}{\begin{eqnarray*}}
\newcommand{\eean}{\end{eqnarray*}}
\newcommand{\dd}{\partial}

\newcommand{\id}{{\rm 1\kern -2.5pt I}} 
\newcommand{\bn}{{\hat n}}

\newcommand{\bx}{{\mathbf x}}

\newcommand{\vl}[0]{\ensuremath{\boldsymbol{\ell}}}
\newcommand{\av}[1]{\left\langle #1 \right\rangle}
\newcommand{\bL}{{\mathbf{L}}}
\renewcommand{\emph}[1]{\textit{#1}}

\newcommand{\BK}[0]{ {BICEP/Keck}\xspace}

\begin{document}

\title{Detecting rotation from lensing in the CMB}

\author[a]{Julien Carron,}
\author[a]{Enea Di Dio}
\author[a]{Ruth Durrer}

\affiliation{D\'epartement de Physique Th\'eorique and Center for Astroparticle Physics,\\
Universit\'e de Gen\`eve, 24 quai Ernest  Ansermet, 1211 Gen\`eve 4, Switzerland}

\date{\today}

\emailAdd{julien.carron@unige.ch}
\emailAdd{enea.didio@unige.ch}
\emailAdd{ruth.durrer@unige.ch}

\abstract{
An excellent estimate of the lensing signal is expected from the availability of deep and high-resolution polarization data in the near future. This is most important to allow for efficient delensing, needed to detect the primordial 
B-mode power and with it the famous tensor-to-scalar ratio.
 Here we discuss in a joint manner estimators of the rotation of polarization, of the second order lensing field rotation, and standard gradient lensing reconstruction. All are most efficient when able to probe the $EB$ power created locally, have comparable reconstruction noise in this regime, and can benefit substantially from delensing. We discuss several ongoing and planned CMB experiments. We determine their noise for lensing field rotation and polarization rotation and discuss their prospects for measuring these effects. 
There is an on-going controversy on whether the lensing field rotation also rotates the polarization -- if so this will be observed at high significance soon with already on going observations of the South Pole Telescope, SPT-3G, in cross-correlation with tracers of large scale structure, as we show in this paper. 
}

\maketitle
\tableofcontents
\section{Introduction}
The cosmic microwave background (CMB) is the most precise cosmological dataset. It has let to the determination of the basic parameters in our Universe with percent and sub-percent precision~\cite{Planck:2018nkj,Planck:2018lbu,Planck:2018vyg} . With present (SPT-3G~\cite{SPT-3G:2014dbx, SPT-3G:2024qkd}, ACT~\cite{ACT:2020gnv, ACT:2023kun} e.g.)
and near future observations of the CMB, especially with the Simon's Observatory, SO~\cite{SimonsObservatory:2018koc}, 
but also with LiteBird~\cite{LiteBIRD:2022cnt,LiteBIRD:2023iiy} and ultimately with the CMB-S4 experiment~\cite{Abazajian:2019eic, 2022ApJ...926...54A}, especially its deepest patch, CMB-S4-deep, the precision of CMB observations will 
be significantly boosted. In particular, the CMB polarization will be mapped to much higher precision on both large and small angular scales. By 2026, observations of SPT-3G on its main survey should reach enough co-added sensitivity to resolve each and every $\Lambda$CDM $B$-mode in this area to $\ell \sim 1100$ with signal to noise larger than unity. On-going design efforts for CMB-S4 plan sensitivities in the deepest areas still higher than that by a large factor, resolving the $B$-mode map to $\ell \sim 2500$ and more. Such a deep data set will not only allow us to put tight constraints on the tensor to scalar ratio $r$ to the small value of about $5\cdot 10^{-4}$ but also to measure the lensing potential spectrum and other signals with very high precision.

In this paper we study whether a new second order lensing effect can be measured: at second order, the coupling of weak gravitational lensing at different redshifts generates not only a contribution to the scalar lensing potential, but the deflection angle also acquires a curl component~\cite{Pratten:2016dsm,Marozzi:2016uob}.
We investigate whether this curl component can be measured either with future CMB experiments alone or when combining CMB experiments with tracers of the large scale structure (LSS) of the Universe.

In the literature there is a controversy whether this curl component leads to a significant rotation of the polarization direction or not. While Refs.~\cite{Lewis:2016tuj,Lewis:2017ans,Namikawa:2021obu} argue that there is (nearly) no rotation (proportional to the deflection angle squared, hence negligible), Refs.~\cite{Marozzi:2016und,Marozzi:2016qxl,DiDio:2019rfy} argue to the contrary (proportional to the shear squared). In this paper we do not enter in this controversy, but we shall show that observations can actually decide about it by comparing the effect of the lensing angle curl on the CMB spectra with and without polarization rotation.

We shall consider the lensing deflection angle {\boldmath$\alpha$} with its curl component $\om = -\frac 12\ep^{ab}\nabla_a\alpha_b$ and a
polarization rotation angle $\beta$. For the theoretical developments we keep $\beta$ arbitrary but we shall consider numerical results setting $\beta = -\om$ as claimed in Refs.~\cite{Marozzi:2016und,Marozzi:2016qxl,DiDio:2019rfy}.
In this case, in the signal to noise ratios we calculate, $\beta$ is therefore not an additional parameter but it is determined by $\omega$.
We shall find that in this case, the chances to measure the
lensing curl are much better than without the polarization rotation effect: Adding polarization rotation, CMB-S4 deep alone will measure $\om$ with a signal-to-noise (SNR) of about 3.9. If we include a template for $\om$ from large scale structure observation, this SNR can be boosted to about 35 and even SPT-3G after its nominal observation time (2019-2026, labeled as SPT-3G-7y in what follows) will achieve SNR $\sim 7$.

This paper is structured as follows: In the next section we introduce the effect of first and second order lensing on the CMB power spectra without and with rotation of the polarization.
In Section~\ref{s:quad} we derive a quadratic estimator for the lensing curl from the CMB alone and determine its signal-to-noise (SNR) for different CMB experiments. We concentrate especially on the difference of whether polarization is rotated or not.
In Section~\ref{s:det} we discuss the prospects for detection of lensing rotation, especially in Subsection~\ref{s:lss} we study the improvement that can be gained when a lensing curl template from an LSS tracer is used.
In Section~\ref{s:con} we discuss our results and conclude.

Assuming that there is no polarization rotation, the prospects of measuring the curl of the deflection angle have already been studied in~\cite{Robertson:2023xkg}. For easy comparison with these results, we shall use the same LSS tracers that have been used in this pioneering paper.
This also helps to better study the effect of a possible rotation of the polarization.
\vspace{1cm}\\
{\bf Notation:}\\
We consider a flat Friedmann Universe with scalar perturbations. The linearly perturbed metric in longitudinal gauge is given by
\be
ds^2= a^2(t)\left[-(1+2A)dt^2+(1-2B)\de_{ij}dx^idx^j\right] \,.
\ee
The coordinate $t$ denotes conformal time.
Here $A(\bx,t)$ and $B(\bx,t)$ are the Bardeen potentials and
\be\label{e:Weyl}
\Phi(\bx,t) = \frac{1}{2}[A(\bx,t)+B(\bx,t)]
\ee
is the Weyl potential that enters the geodesic motion of photons.
An overdot denotes the derivative w.r.t. conformal time such that $\dot a/a = \HH$ is the conformal Hubble parameter while $\dot a/a^2=H$ is the physical Hubble parameter. We work in units such that the speed of light is unity.

Since the sign conventions for lensing vary in the literature which can lead to confusion, let us give our conventions here in some detail.
We first write them in complex notation for tensor fields on the sphere, where $\eth^{\pm}$ denotes the spin raising and lowering operator, see e.g.~\cite{Durrer:2020fza}
\begin{align}
	\alpha(\bn) &= - \eth^+(\psi(\bn)  + i \Omega(\bn) ) \\ 
	\kappa(\bn) + i\omega(\bn) &= \frac 12 \eth^-\alpha(\bn) \\ 
	\gamma_1(\bn)  + i \gamma_2(\bn) &= \frac 12 \eth^+\alpha(\bn)\,. 
\end{align}
Here $\psi$ is the lensing potential and $\Omega$ is the curl potential
for the deflection angle $\al=\al_\vth+i\al_\varphi$ that is defined via
\be
\bn' = \bn+ {\boldsymbol\al}\,, 
\ee
where $\bn$ is the (lensed) observation direction and $\bn'$ is the unlensed direction.
In the flat sky approximation, we may use
\be
\eth^+ =-(\dd_1+i\dd_2) \qquad \eth^- =-(\dd_1-i\dd_2).
\ee
This can be thought of placing ourselves at the equator, with the first direction pointing south along $\boldsymbol{e}_\vth$, the second east along $\boldsymbol{e}_{\varphi}$, and neglecting sky curvature.
The flat sky relations between the above quantities then become
\begin{align}
	\boldsymbol{\alpha}(\bn)&= \bnabla \psi(\bn) - \bnabla \times \Omega(\bn) = \begin{pmatrix}
		\partial_1 \psi - \partial_2 \Omega \\ \partial_2 \psi + \partial_1 \Omega
	\end{pmatrix}\\
	\kappa(\bn) &= -\frac 12 \bnabla \cdot \boldsymbol{\alpha}(\bn) = -\frac 12 \partial_i \alpha_i(\bn)  = - \frac 12 \Delta \psi\\ 
	\omega(\bn) &= -\frac 12 \bnabla \times \boldsymbol{\alpha}(\bn)  = -\frac 12\epsilon^{ij}\partial_i \alpha_j(\bn) =  -\frac 12 \Delta \Omega \\ 
	\gamma_1(\bn) &= -\frac 12 (\partial_1 \alpha_1(\bn) - \partial_2 \alpha_2(\bn) ) \\
	\gamma_2(\bn) &= -\frac 12 (\partial_1 \alpha_2(\bn) + \partial_2 \alpha_1(\bn) ) \\
 \left(\ep_{ij}\right) &= \begin{pmatrix}
		0 & 1 \\ - 1 & 0
	\end{pmatrix} \,.
\end{align}
Note that the 2d curl of the (pseudo) scalar  $\Om$ can be obtained from the 3d curl assuming that  $\Om$ represents a vector in $3$-direction or simply via $(\bnabla \times \Omega)_i =\ep_{ij}\dd_j\Om$ . 
These conventions give the magnification matrix
\begin{equation}\label{e:magmat}
	\left(\delta_{ij} + \frac{\partial \alpha_i}{\partial n_j}\right) = \begin{pmatrix}
		1- \kappa  - \gamma_1 & -\gamma_2 + \omega \\ - \gamma_2 - \omega & 1-\kappa + \gamma_1
	\end{pmatrix} \,.
\end{equation}
These conventions imply that in the presence of a positive $\omega$, an image appears to the observer rotated {\it clockwise} by this angle.\footnote{Careful: seen from inside the sphere, the standard polar basis vectors $\boldsymbol{e}_1 = \boldsymbol{e}_{\vth}, \boldsymbol{e}_2 = \boldsymbol{e}_{\varphi} $ forms a left-handed coordinate system.} This is the opposite sign (of $\omega$ and $\Omega$) to that adopted by the pioneering paper~\cite{Hirata:2003ka} or the well-known CMB lensing review~\cite{Lewis:2006fu}, as well as many subsequent works in the flat-sky approximation.
Our conventions are, however, consistent for example to {\it Planck} and to widespread CMB software~\cite{Gorski:2004by, Zonca2019}, where the gradient (G) and curl (C) modes of a vector field are expanded in vector spherical harmonics as
\begin{equation}
	\alpha(\bn) = - \sum_{\ell  m}(G_{\ell m} + iC_{\ell m}) {\,}_{1}Y_{\ell m}(\bn),
\end{equation}
where in our case $G_{\ell m} = \sqrt{\ell (\ell  + 1)} \psi_{\ell m}$ and $C_{\ell m} = \sqrt{\ell (\ell  + 1)} \Omega_{\ell m}$.
Also the polarization, which is parallel transported along the light ray can get rotated due to a rotation of the local reference frame on the screen. The complex polarization, $P(\bn)=Q(\bn)+iU(\bn)$ given in terms of the Stokes parameters $Q$ and $U$, is a spin-2 field. A rotation by an angle $\beta(\bn)$ rotates the polarization into
$P(\bn) \ra e^{-2i\beta(\bn)}P(\bn)$. In our conventions, opposite to the IAU polarization conventions owing to $\boldsymbol{e}_{1} = \boldsymbol{e}_{\theta}$ pointing south instead of north, a positive $\beta$ is seen as a \emph{counter-clockwise} rotation for the observer.
 
\section{CMB lensing at second order}\label{s:lens}
In the presence of purely scalar perturbation, the foreground gravitational field deflects CMB photons. At first order the deflection angle is given by (see, e.g.~\cite{Lewis:2006fu})
\be\label{e:alpha1}
\al^{(1)}_i(\bn, z_*) = \nabla_i\psi(\bn)= -2\nabla_i\hspace{-1mm}\int_0^{r_*}\hspace{-3mm}dr\frac{r_*-r}{r_*r}\Phi(r\bn,t_0-r)\,.
\ee
Here $\psi(\bn)$ is the lensing potential for the CMB, $r_*=r(z_*)$ is the comoving distance out to the CMB redshift $z_*$ and $\Phi(\bx,t)$ is the Weyl potential. The derivative $\nabla_i$ is a 2d derivative on the sphere.

At first order, Eq.~\eqref{e:alpha1}, the deflection angle is determined along the unperturbed, background geodesic, $(t_0-r, \bn r)$, the Born approximation, while at the next order, the first order deflection of the path due to foreground lensing is included,
\bea \label{eq:alpha2}
\al^{(2)}_i(\bn, z_*)   &=& -2\int_0^{r_*}dr\frac{r_*-r}{r_*r}\nabla_j\nabla_i\Phi(r\bn,t_0-r)\al^{(1)}_j(\bn, z(r)) \nonumber \\
&=& 4\int_0^{r_*}dr\frac{r_*-r}{r_*r}\nabla^j\nabla^i\Phi(r\bn,t_0-r)\int_0^{r}dr'\frac{r-r'}{rr'}\nabla_j\Phi(r'\bn,t_0-r') \,.
\eea 
On the second line we have introduced the lowest post-Born contribution for the deflection angle. There are also other second order contributions to the deflection angle, but the post-Born effect dominates by far, see~\cite{Marozzi:2016qxl}. This is the approximation we shall use in this paper.
Due to contributions from different distances, $r\neq r'$ the second order term contributes also a curl component to the deflection angle, see~\cite{Fanizza_2013,Lewis:2016tuj,Marozzi:2016und}
\be \label{eq:omega}
\om(\bn) = -\frac{1}{2}\ep^{ij}\nabla_i\al_j =
2\ep^{ij}\int_0^{r_*}dr\frac{r_*-r}{r_*r}\nabla_i\nabla^k\Phi(r\bn,t_0-r)\int_0^{r}dr'\frac{r-r'}{rr'}\nabla_j\nabla_k\Phi(r'\bn,t_0-r') \,.
\ee
Note that in two dimensions the curl is a pseudo-scalar quantity.

\begin{figure}[!t]
\centering
	\includegraphics[width=0.6\columnwidth]{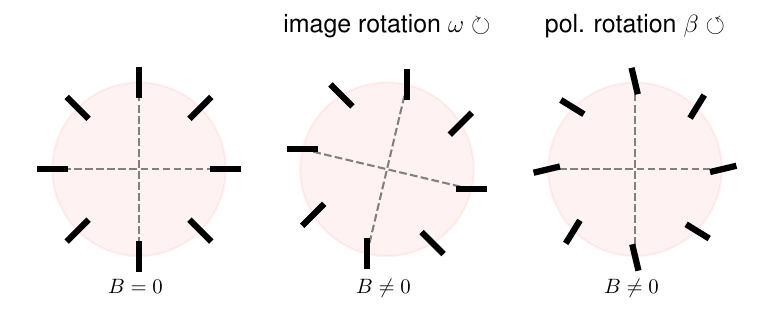}
	\caption{\label{fig:bbcross_sign}A parity-invariant, pure $E$-mode pattern, after a clockwise image rotation by a small (positive) angle $\omega$, keeping the polarization directions at each point fixed (center panel), or after a counter-clockwise polarization rotation by the same small angle $\beta$, keeping the points of the image fixed (right panel). The curliness ($B$-mode) of the resulting patterns is clearly of the same sign.  Hence a constant clockwise image rotation generates the same $B$-polarization as a constant counter-clockwise polarization rotation. Large-scale magnification never creates any $B$-modes but only affects the $E$-mode. The case of large shearing lenses is discussed in appendix~\ref{a:squeeze} .}
\end{figure}

In Refs.~\cite{Lewis:2016tuj,Lewis:2017ans} it is argued that while this curl does lead to a rotation of images, i.e.~it affects weak lensing, it does not influence the polarization. In Refs.~\cite{Marozzi:2016und,Marozzi:2016qxl,DiDio:2019rfy} it is argued to the contrary that a rotation of polarization can only be defined w.r.t.~directions of images and hence this rotation also affects polarization, rotating it by an angle\footnote{While in Ref~\cite{Marozzi:2016qxl} the polarization rotation angle was defined as it is here, in Ref.~\cite{DiDio:2019rfy} it is defined with the opposite sign which leads to $\beta=\om$.
} $\beta=-\om$. An illustration how image rotation by a constant angle $\om$ is equivalent to polarization rotation by $\beta=-\om$ is illustrated in Fig.~\ref{fig:bbcross_sign}. We do not want to take a stand here, but we simply study whether such a rotation, if present in the polarization could be detected.
The lower black line in Fig.~\ref{fig:N0comp} shows the angular power spectrum, $C_L^\om$, of $\om$ as the lower black line (for standard Planck cosmological parameters). It peaks at $L\sim 200$ and its maximum is about 5 orders of magnitude smaller than the one of the convergence power spectrum, $C_L^\kappa$ which is shown as the upper black line in the figure. 

\section{Quadratic estimators for polarization rotation and lensing curl}\label{s:quad}

\subsection{Generalities}
In this section we derive a quadratic estimator (QE) for the rotation $\beta$ of the polarization.  We then study how well this can be measured with near future CMB experiment if $\beta=-\om$. Quadratic estimators for the CMB lensing potential have been introduced in~\cite{Hu:2001kj,Okamoto:2003zw}. More recent developments can be found in~\cite{Maniyar:2021msb}.  A quadratic estimator for polarization rotation was first derived in \cite{Yadav:2009eb} within the flat sky approximation. In \cite{Gluscevic:2009mm} this is generalized to the curved sky.  For completeness, we discuss here how these estimators are obtained. In this paper we use the `General Minimum Variance' (GMV) estimators~\cite{Maniyar:2021msb}, which are more optimal than the original ones by up to 10\% in the case of lensing, but much less than that for polarization rotation.

Optimally-weighted quadratic estimators are often most concisely written in configuration space. This is also the case for polarization rotation. The quickest way towards the result is to make use of the general fact that the optimal estimator (optimal in the statistical sense) of a source of anisotropy is always given by the gradient of the likelihood with respect to that source\footnote{This is a very general result that holds for the reconstruction of any parameter, which can be used to devise efficient data compression schemes~\cite{Carron:2013nfa, Alsing:2017var}: By construction, likelihood gradients have minimum variance, because they saturate the Fisher information content of the data (this is in fact how Fisher information is defined).}. Since the effect of polarization rotation is particularly simple in configuration space, construction of the likelihood gradient is simple, and the only task remaining is to calculate the response of the estimator to the source, which is then used to normalize the gradient. Furthermore, in harmonic space optimal quadratic estimator weights are given by the bi\-spec\-trum between the source of interest and the fields used to reconstruct it. This is reviewed for the case of lensing in~\cite[Appendix C]{Carron:2024mki}, and we follow their notation here.
\newcommand{\thickbar}[1]{\accentset{\rule{.5em}{1pt}}{#1}}
According to our conventions, the effect of a position dependent rotation on the CMB Stokes intensity and polarization fields, which have spin $s=0$ or $s=2$, is
\begin{equation}
	{}_sX(\bn) \rightarrow  e^{- i s \beta(\bn)}{}_sX(\bn).
\end{equation}
Hence, the gradient of the CMB map with respect to $\beta(\bn)$ is simply $-is \:{}_sX(\bn)$. The CMB likelihood weighs pairs of observed CMB maps by the inverse data covariance $C^{\text{dat}}$. We can define the inverse-covariance weighted maps in harmonic space as 
\begin{equation}\label{eq:ivf}
	\bar X_{\ell m} = \left[C^{\text{dat}}_\ell\right]^{-1}_{XY} Y^{\rm dat}_{\ell m}
\end{equation}
where in harmonic space $X = T, E, B$. Further, the `Wiener-filtered' CMB maps are given by
\begin{align}
	T^{\rm WF}_{\ell m} = C^{TT}_\ell\bar T_{\ell m} +  C^{TE}_\ell\bar E_{\ell m}, 
\quad	 E^{\rm WF}_{\ell m} = C^{EE}_\ell\bar E_{\ell m} +  C^{TE}_\ell\bar T_{\ell m}, \label{eq:EWF}
	 \quad  B^{\rm WF}_{\ell m} = 0.
\end{align}
The likelihood gradient (our quadratic estimator for $\beta$) becomes
\begin{align}\label{eq:polrotreal}
	\bar \beta(\bn) &=\frac 12 \sum_{ s =\pm 2} \: {}_{-s}\bar{X}(\bn)\: (-i s) {}_{s}X^{\rm WF}(\bn) 
 \equiv \sum_{LM} \bar \beta_{LM} \:{}_{}Y_{LM}(\bn) \, ,
 \end{align}
where
 \be
 {}_{2}\bar{X}(\hat n) =-\sum_{\ell m} (\bar E_{\ell m} +i \bar B_{\ell m} ){}_{2}Y_{\ell m}(\hat n)\,, \quad 
  {}_{-2}\bar{X}(\hat n) =  -\sum_{\ell m}(\bar E_{\ell m} -i \bar B_{\ell m}) {}_{-2}Y_{\ell m}(\hat n)\,.
 \ee

The harmonic space weights can be obtained directly transforming the first equation to harmonic space. Doing so, it is convenient to introduce
\begin{equation}
	{}_{s}F_{\ell_1\ell_2L}^{\pm} = \frac 12 \left[\begin{pmatrix}
		\ell_1 & \ell_2 & L \\-s& s & 0
	\end{pmatrix} \pm \begin{pmatrix}
		\ell_1 & \ell_2 & L \\ s & -s & 0
	\end{pmatrix} \right]\,,
\quad
\mbox{ where } ~
\begin{pmatrix}
		\ell_1 & \ell_2 & \ell_3 \\ m_1 & m_2 & m_3
	\end{pmatrix}
\end{equation}
    are the Wigner 3j symbols.
These factors contain the relevant `trigonometric' weighting: $_{s}F^+$ has even parity, and includes a cosine of $s$ times the angle between $\ell_1$ and $\ell_2$ in the flat-sky limit~\cite{Carron:2024mki}, and $_{s}F^-$ has odd parity, including a sine of that angle. One obtains
\begin{align}
	\bar \beta_{LM} =& \frac{(-1)^M}{2} \sum_{\ell_1 m_2 \ell_2m_2}\begin{pmatrix}
		\ell_1 & \ell_2 & L \\ m_1 & m_2 & -M
	\end{pmatrix} 
 \sqrt{\frac{(2\ell_1 + 1)(2\ell_2 + 1)(2L + 1)}{4\pi}} \nonumber  \\
	&\times (-4) \left[ {}_{2}F_{\ell_1\ell_2L}^+ \bar B_{\ell_1 m_1}+ i \:{}_{2}F_{\ell_1\ell_2L}^-  \bar E_{\ell_1 m_1} \right]E^{\text{WF}}_{\ell_2 m_2} \, .
\end{align}
The Wiener-filtered $E$-mode is given by~\eqref{eq:EWF}.
These weights can also be obtained by a brute-force minimum variance calculation for the optimal weights. For practical purposes, the position-space version, Eq.~\eqref{eq:polrotreal}, is more straightforward, since allowing evaluation of $\bar \beta$ with just a few harmonic transforms.

The lensing estimators (likelihood gradients with respect to the deflection vector field) are given in very similar notation by~\cite{Aghanim:2018oex}
\begin{align}
	{}_{1}\bar \alpha(\bn) &=-\sum_{ s =0, \pm 2} \: \frac{ _{-s}\bar{X}(\bn)}{2 - \delta_{s0}}\:  \eth^+ {}_{s}X^{\rm WF}(\bn) 
 \equiv - \sum_{LM} \left(  \frac{\bar \psi_{LM} + i \bar \Omega_{LM}}{\sqrt{L(L + 1)}}\right)\:{}_{1}Y_{LM}(\bn) \, .
\end{align}
See~\cite{Namikawa:2011cs} or \cite[appendix C]{Carron:2024mki} for the lensing curl harmonic space weights.
The gradients with respect to $\kappa$ and $\omega$ relate to those of $\psi$ and $\Omega$ according to the chain rule and the definitions $\kappa = - \Delta \psi/2$, $\omega= - \Delta\Omega/2$,
\bea
	\bar \kappa_{LM} &=& \frac{2 }{L(L + 1)}\bar \psi_{LM} \,,\\
\bar \omega_{LM} &=& \frac{2 }{L(L + 1)}\bar \Omega_{LM} \,. 
\eea
For a signal entering both fields $\beta$ and $\omega$, the optimal estimator for that signal can again be computed by the chain rule as the corresponding weighted sum of $\bar \beta_{LM}$  and $\bar \omega_{LM}$.

\subsection{Squeezed limits and B-mode powers}\label{ss:squeeze}
\begin{figure}[!t]
\centering
	\includegraphics[width=0.99\columnwidth]{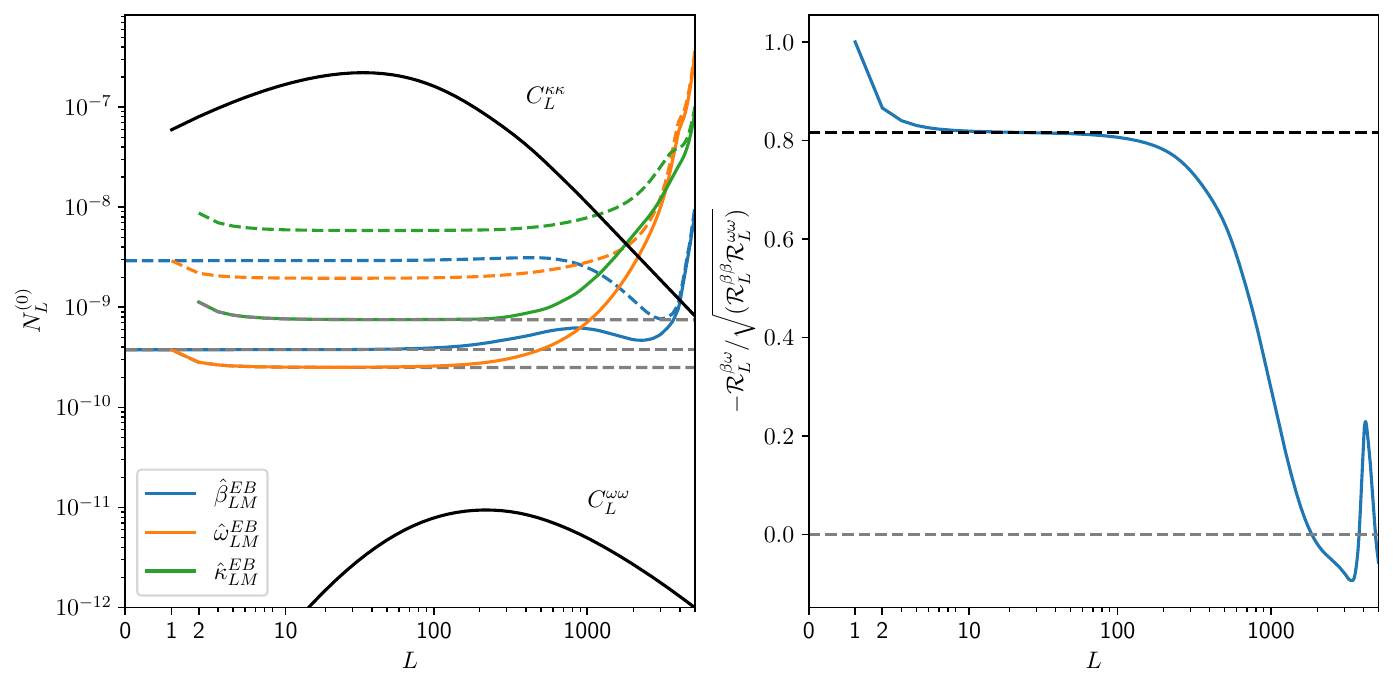}
	\caption{\label{fig:N0comp}\emph{Left panel:} Map-level reconstruction noise  $N^{(0)}_L$ for the polarization rotation estimator ($\hat \beta$ blue) and lensing curl mode estimator ($\hat \omega$, orange), for a configuration approaching that of the CMB-S4 deep configuration. This configuration can provide the lowest reconstruction noise owing to its ability to delens almost 90\% of the dominant gradient lensing $B$-power on the relevant scales (results without delensing are shown as the dashed lines). Also shown for comparison is the case of the lensing gradient mode ($\hat \kappa$, green). The lensing curves have a slight scale-dependence at low $L$ owing to information from the shear.   Grey dashed lines show the analytical low-$L$ behavior as discussed in details in the main text. Solid black are the $\kappa$ spectrum and our fiducial post-Born lensing curl spectrum.\\
	\emph{Right panel:} (Negative of) the cross-correlation coefficient of the $EB$ polarization and lensing rotation estimators (they are of the same parity and signal will leak from one to the other.) The degeneracy for the dipole is perfect and broken for $L \ge 2$ by the information from the lensing shear curl-mode.} 
\end{figure}

\begin{figure}
	\includegraphics[width=1\columnwidth]{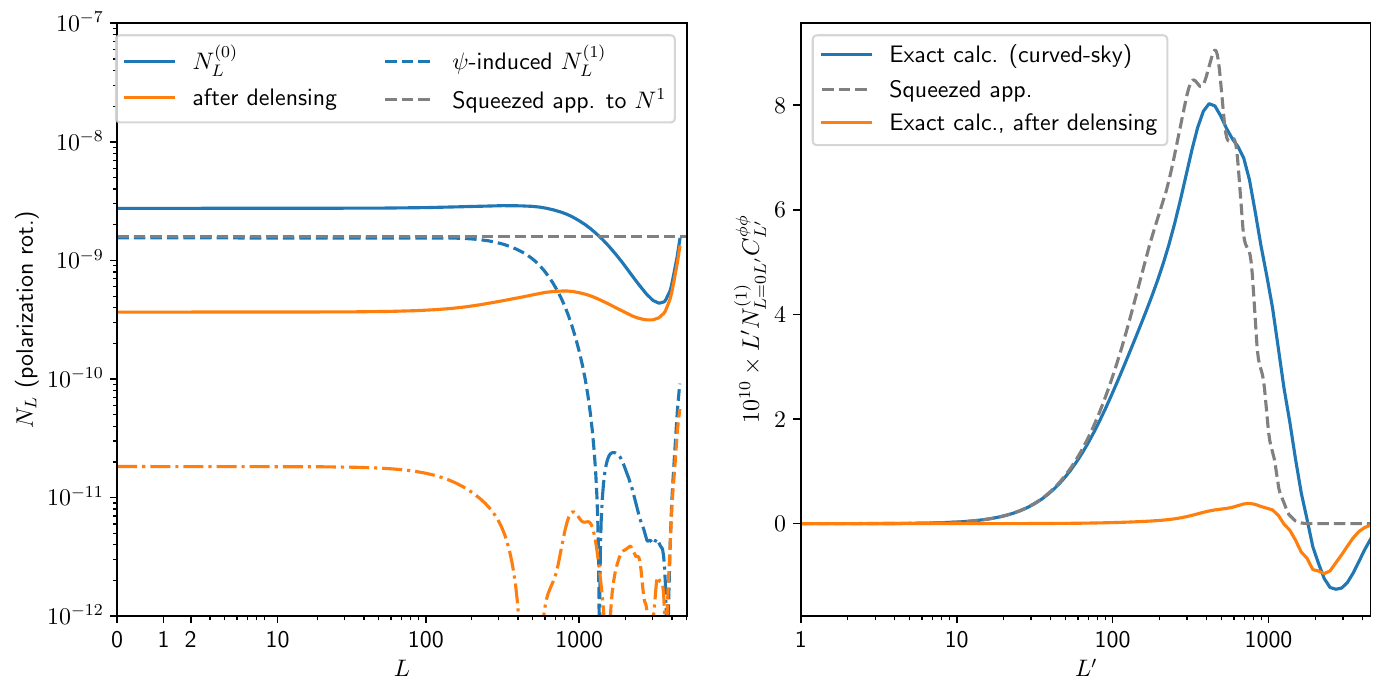}
	\caption{\label{fig:N1comp}\emph{Left panel:} Lensing potential $\psi$-induced $N_L^{(1)}$  bias, 
 calculated using the flat-sky approximation, entering the spectrum of the polarization rotation ($\beta$) reconstruction, before (blue, the quadratic estimator case) and after  delensing (orange, dot-dashed shows negative values) and in comparison to the leading Gaussian $N_L^{(0)}$ biases (solid lines). Dashed grey line is the curved-sky analytic approximation to the large-scale $N^{(1)}$ obtained in the appendix.
This is for the deepest configuration tested in this work. 
	In the QE case, $N_L^{(1)}$ is a very significant contribution on large scales, half the size of the leading bias. Delensing is highly efficient at removing this bias, by two orders of magnitude. $N_L^{(1)}$ biases induced by the lensing curl $\omega$ or $\beta$ itself are much smaller and not shown here.
	The $\psi$-induced lensing curl $N_L^{(1)}$ bias behaves very similarly. \emph{Right panel:} Contributions of lenses $L'$ to the large scale $N_{L\sim 0}^{(1)}$ of the left panel. The blue line shows the exact result in curved-sky geometry, and orange the same after delensing. The dashed grey the squeezed approximation. This shows that the very good agreement on the left panel is to some part coincidental.}
\end{figure}
The lensing post-Born curl mode peaks at $L\sim 200$, and many sources of interest for polarization rotation are relevant on even larger scales. For example, a constant bi-refringence angle enters the monopole $L=0$ of the polarization rotation estimator. Hence, it is worthwhile discussing the squeezed limits of these estimators and of their (co-)variance. We write generally $\mathcal R^{xy}_L$ for the response of the estimator optimized for anisotropy source $x$ to a possibly different anisotropy source $y$.
The leading (Gaussian) reconstruction noise (usually called $N_L^{(0)}$) of these estimators is equal to the inverse response
\begin{equation}
N^{xx(0)}_L = \frac{1}{\mathcal R^{xx}_L}
\end{equation}
We defer to Appendix~\ref{a:squeeze} for some details of this section.

On large angular scales (small $L$), quadratic estimators measure the change in the local isotropic power spectrum ($\av{X_{\vl_1}X_{\vl_2}}$ with $-\vl_1 \sim \vl_2$) caused by the long mode of the anisotropy source. 
The long mode can be treated as constant locally, and the spectrum response can be obtained simply through the flat-sky approximation.

The linear response of the $C_\ell^{EB}$ spectrum to a rotation $\beta$ is $-2 \beta C_\ell^{EE}$ and that of $C_\ell^{TB}$ is $-2\beta C_\ell^{TE}$. Hence (see Appendix~\ref{a:squeeze}),  the estimator responses at low $L$ will be given by the Fisher-like trace (up to $(L/\ell)^2$ corrections). Neglecting any primordial $BB$ spectrum and non-perturbative corrections we obtain
\begin{align}\label{eq:sqz}
	&\mathcal R_L^{\beta\beta, EB} \sim  \sum_\ell\frac{2\ell + 1}{4\pi} \frac{-2C_\ell^{EE}}{C_\ell^{EE} + N_\ell^{EE}}\frac{-2C_\ell^{EE}}{C_\ell^{BB} + N_\ell^{BB}} \, , \\
		&\mathcal R_L^{\beta\beta, TB} \sim  \sum_\ell\frac{2\ell + 1}{4\pi} \frac{-2C_\ell^{TE}}{C_\ell^{TT} + N_\ell^{TT}}\frac{-2C_\ell^{TE}}{C_\ell^{BB} + N_\ell^{BB}} \, .
\end{align} 
The summands are the contribution of each local small scale mode to the large scale result.
The $TB$ estimator is not competitive compared to $EB$; for deep polarization experiments these equations show that the $TB$ response is smaller by a factor $\rho_\ell^2 \sim 1/10$ compared to $EB$, where $\rho_\ell= C_\ell^{TE}/\sqrt{C_\ell^{TT}C_\ell^{EE}}$ is the small scale cross-correlation coefficient between $T$ and $E$. 
As soon as $N_\ell^{BB}$ goes below the lensing $B$-power, it is clear that delensing can bring significant gains, with the reconstruction noise scaling linearly with the residual lensing power in this regime.

On the other hand, a locally constant rotation, as it has odd parity, does not change $C_\ell^{EE}$ nor $C_\ell^{TE}$, so that $TE$ and $EE$ will not provide any information on large scale modes (we have at low $L$ $\mathcal R_L^{\beta\beta, TE} \sim	\mathcal O(L/\ell)^2 \quad	\mathcal R_L^{\beta\beta, EE}\sim \mathcal O(L/\ell)^2)$; $EB$ is the only relevant piece for large-scale polarization rotation reconstruction.

Now, a locally constant lensing rotation mode $\om$  has the same effect on the local CMB power spectra as a rotation of polarization (in particular the change in $C_\ell^{EB}$ is $-2\omega C_\ell^{EE}$): so that there will be degenaracies between the lensing curl and polarization rotation estimators. This degeneracy will be broken (for $L > 1$) to some degree by the information provided by the local lensing shear $B$-mode sourced by the global $\omega$.

The responses for the lensing curl mode estimator will therefore look similar. We have in fact (again, up to $(L/\ell)^2$ corrections)
\begin{align}
	\mathcal R_L^{\omega\omega, EB}  \sim \mathcal  R_L^{\beta\beta, EB} \times
	\left( 1 + \frac 12 \frac{(L - 1)(L + 2)}{L (L + 1)} \right).
\end{align}
The second factor in the parenthesis is the information coming from the shear, and gives some scale-dependence at the very lowest multipoles. This factor can be understood as follows: $(L-1)(L + 2)/L(L + 1)$ is the factor relating the harmonic modes of the curved sky shear $B$ term ($ \gamma_1 + i \gamma_2 = -\frac{ i}2 \eth^+\eth^+\Omega$)  to that of $\omega = -\frac 12 \eth^-\eth^+ \Omega$. Furthermore, there is a factor $1/2$ reduction of the response (or factor of $2$ increase in noise), since there are now two independent shear components. The response of the lensing curl is larger by 3/2  at sufficiently large $L$ but still respecting the squeezed limit.

In Fig.~\ref{fig:N0comp} we show reconstruction noise curves together with the limiting behavior just discussed, for the deepest CMB configuration we consider in the this paper, CMB-S4 deep, as described in Appendix~\ref{ap:exp}. 
The delensing is predicted following the spectrum-level analytic iterative procedure first proposed for the $EB$-estimator in \cite{Smith:2010gu}: starting from the standard lensing quadratic estimator noise level, one assumes the corresponding amount of lensing power can effectively be removed from the map, providing new maps with less lensing power.
One then iterates this calculation until convergence. 
This procedure and other variants using the full polarized data has been shown to match accurately detailed map-level delensing work on simulated CMB-S4 maps~\cite{CMB-S4:2020lpa, Belkner:2023duz}. 
Also shown for comparison are the noise curves for the dominant lensing gradient mode. In this case, since a locally constant magnification does not generate any $EB$ signal, the response is that of the shear-piece only,
\begin{equation}
	\mathcal R_L^{\kappa \kappa, EB} \sim \mathcal  R_L^{\beta\beta, EB}  \frac 12 \frac{(L - 1)(L + 2)}{L (L + 1)}
\end{equation}
with  $\mathcal R_L^{\beta\beta, EB}$ from Eq.\eqref{eq:sqz}.

As already advertised, a $\beta$ or $\omega$ signal will leak into both estimators. The low-$L$ cross-responses are identical to \eqref{eq:sqz},
\begin{equation}
		\mathcal R_L^{\omega\beta, EB} \sim \mathcal R_L^{\beta \omega, EB} \sim  \mathcal  R_L^{\beta\beta, EB}.
\end{equation}
The cross-correlation coefficient of the two estimators is shown on the right panel of Fig.~\ref{fig:N0comp}. The dashed line indicates $\sqrt{2/3} \sim 0.82$, obtained in the squeezed limit but away from the dipole using our results from this section. On the other hand, from parity arguments the two estimators have no linear responses to $\kappa$.

Let us briefly discuss limits on polarization rotation by a constant angle $\bar\beta$. Let an adequate survey window be $W(\bn)$. After quadratic or iterative reconstruction of the polarization rotation map $\hat \beta(\bn)$, the monopole may be inferred from the average
\begin{align}
	\hat {\bar\beta} &= \frac 1 {A} \int d^2 \hat n\: \hat \beta(\bn) W(\bn) 
	= \frac 1 A \sum_{LM} \hat\beta_{LM} W^*_{LM}
\end{align}
where $A = \int d^2n W(\bn)$ is the effective area of the patch used to infer the monopole.
The reconstruction noise variance of $\hat{\beta}_{LM}$ is approximately given by $N_L^{\beta\beta(0)}$. Taking the variance of this equation, one obtains in the absence of a signal
\begin{equation}
	\sigma^2_{\bar \beta}\equiv \av{ \left(\hat {\bar\beta} -\bar \beta  \right)^2} = \frac 1 {A^2}\sum_{L} N_L^{\beta\beta(0)} \sum_{M=-L}^L |W_{LM}|^2.
\end{equation}
Since $N_L^{(0)}$ is very flat on large scales where the window is relevant (see Fig.~\ref{fig:N0comp}), we can take $N^{(0)}_L$ out of the $L$ sum. The sum of the squared window harmonics can then be written as the sky integral of the squared window. The result then simplifies to
\begin{equation}
	\sigma^2_{\bar \beta}	 \sim \frac{N^{\beta \beta(0)}_{L\sim0}}{4\pi f_{\rm sky}}, 
\end{equation}
where $1 / (4\pi f_{\rm sky})$ is short-hand notation for $\int d^2 \hat n \:W^2(\bn) / (\int d^2 \hat n\: W(\bn))^2 $ (for a tophat window this simply becomes $1/\int d^2\hat n\: W(\bn)=1/A=1/4\pi f_{\rm sky}$).
The standard deviations $\sigma_{\bar \beta}$ corresponding to the configurations looked at in this work are listed in the last column of Table I.
As one sees there, e.g.~for S4-deep  a constant polarization rotation angle is detected at $5\si$ if $\bar\beta>0.5$ arcmin. For PICO even $\bar\beta>0.4$ arcmin are detected at $5\si$ and even SPT-3G can announce a $5\si$ detection if $\bar\beta>1.9$ arcmin.

Since $\hat \beta$ has a non-zero response to $\omega$, there is contamination from the guaranteed post-Born field rotation in the monopole angle. The $\hat \beta$ estimator normalization $(1/\mathcal R_L^{\beta\beta})$, cancels perfectly at low-$L$ the cross-response $\mathcal R_L^{\beta \omega} \sim \mathcal R_L^{\beta \beta}$, such that $\omega$ enters just as
\begin{equation} \label{eq:monopole}
	\hat {\bar \beta} \ni \frac 1 A \sum_{LM}\omega_{LM} W^*_{LM}.
\end{equation}
This is of course negligible since $C_L^{\omega\omega}$ is so small compared to $N_L^{\beta \beta(0)}$.
\newcommand{\nlev}{n_{\text{lev}}}
\newcommand{\fsky}{f_{\text{sky}}}

 Let us also qualitatively estimate how delensing reduces the reconstruction noise: the reconstruction noise $N^{(0)}$ scales (in the case of $EB$) with one power of the data total $E$ power and one of the total $B$ power. Seen as a function of the small scale instrumental noise level $\nlev$, we have thus the limiting behaviours
\begin{equation}
	N^{(0)} \propto \nlev^{4} \quad (\text{for } \nlev \gg C^{EE}) \quad\text{while} \quad N^{(0)} \propto \nlev^{0} \quad (\text{for }\nlev \ll C^{BB, \text{lensing}},\text{ no delensing.})
\end{equation} 
In the more relevant intermediate regime, ($C^{EE}\gg \nlev \gg C^{BB})$ the noise is $N^{(0)} \propto \nlev^2$.
Delensing helps moving down somewhat the $E$ threshold by lowering the damping tail, but mostly by removing the $B$ threshold. For very deep experiments, the residual lensing $B$ power scales essentially with the noise level, restoring the scaling
\begin{equation}
		N^{(0)} \propto \nlev^{1} \quad (\text{for } \nlev \ll C^{BB, \text{lensing}}, \text{ with delensing}),
\end{equation}
with in practice still slightly super-linear behavior in the cases of interest in this paper. 

The SNR on the spectrum of noise-dominated fields scales inversely as $N_L^{(0)} /\sqrt{\fsky (2L + 1)}$, with $\fsky (2L + 1)$ the effective number of modes available. Linear scaling of $N_L^{(0)}$ with $\nlev$ implies that for fixed observing time, the error on the spectrum is independent of sky fraction, since $\nlev \propto \sqrt{f_{\rm sky}}$. Hence, the super-linear behaviour of $N^{(0)}$ has the consequence that in order to constrain the spectra of $\beta$ and $\omega$ (or small-scale $\kappa$), it is better to observe the same small patch for longer than observing a larger area.
To constrain a constant birefringence angle, the situation is different: $\sigma_{\bar\beta}$ scales now according to $\sqrt{N^{(0)}/ \fsky}$, making wider experiments generally better (in the absence of any other considerations of course).

The estimator power spectra obtain additional contributions besides $N_L^{(0)}$. The most relevant one is a signal term from secondary trispectra contractions, called $N_L^{(1)}$\cite{Kesden:2003cc, Aghanim:2018oex}, which can be expressed using 6j Wigner symbols~\cite{Hu:2001fa, Hanson:2010rp}. In the case  of standard gradient lensing, these additional biases are typically significantly smaller that the primary signal $C_L^{\psi\psi}$, except on small scales. It is also known that the $N_L^{(1)}$ signal induced by birefringence can affect lensing spectrum reconstruction to some extent~\cite{Cai:2022zad, Cai:2024zau}. However since $C_L^{\psi\psi}$ is so much larger than $\beta$ or $\omega$, the guaranteed lensing-induced $N^{(1)}$ on the rotation and curl estimation are very substantial. We review this bias in appendix~\ref{app:biases}, and obtain there also  a useful handy approximation at low $L$. 
In Fig.~\ref{fig:N1comp}, we show the $\psi$-induced bias for polarization rotation  for the CMB-S4 deep configuration. The blue lines show the QE case, comparing $N^{(0)}$ to $N^{(1)}$. Interestingly,  $N^{(1)}$ is only about a factor 2 smaller than $N^{(0)}$. The calculations of $N^{(1)}$ are performed using the flat-sky expressions from~\cite{Aghanim:2018oex}, but we show as a check in dashed grey the analytic, curved-sky geometry approximation we obtain in the appendix: 
\begin{align} 
	n^{(1)\beta \beta}_{L}&  =\sum_{L'}n^{(1)\beta \beta}_{LL'}  \qquad \mbox{where for small $L$ and $L'$} \nonumber \\
 n^{(1)\beta \beta}_{LL'}  & \sim  \left( \frac{2L' + 1}{4\pi} \right)C_{L'}
	^{\kappa\kappa}\frac 12 \frac{(L'-1)(L' + 2)}{L'(L' +1)} \sum_{2\ell \ge L',L}  \frac{2\ell + 1}{4\pi}\left(\frac{-2C_\ell^{EE}}{C_\ell^{EE} + N_\ell^{EE}}\frac{-2C_\ell^{EE}}{C_\ell^{BB} + N_\ell^{BB}} \right)^2 
.  \label{eq:N1bb}
\end{align}
The first factor is half the variance of the shear field $E$-mode. The second collects the response of the small scale $EB$ power in the same way than the estimator responses, Eq. \eqref{eq:sqz}, and generalizes easily to any quadratic estimator. Note that for large $L'$ this approximation is not valid and $n^{(1)\beta \beta}_{LL'}$ can even become negative, see Fig.~\ref{fig:N1comp} (even $n^{(1)\beta \beta}_{L}$ can become negative, after the broad-band $C^{\kappa\kappa}_{L'}$ has been removed by delensing).
We have also confirmed these predictions by generating full-sky CMB simulations and comparing the power spectrum of the quadratic estimate to the predictions, finding excellent agreement.
The  $\psi$-induced lensing $N^{(1)}$ biases behave in a very similar manner: at low $L$, the $N^{(1)}$ terms of the $EB$ quadratic estimators obey
\begin{align}
	n^{(1)\omega \omega}_{L} &\sim n^{(1)\beta \beta}_{L} \times \left(1+\frac 12 \frac{(L-1)(L + 2)}{L(L + 1)}\right) \\ \nonumber
		n^{(1)\beta \omega}_{L} &\sim n^{(1)\beta \beta}_{L}\\ \nonumber
	n^{(1)\kappa \kappa}_{L} &\sim n^{(1)\beta \beta}_{L}\times \left(\frac 12 \frac{(L-1)(L + 2)}{L(L + 1)}\right)\\ \nonumber
	n^{(1)\kappa \beta}_{L} &\sim n^{(1)\kappa \omega}_{L}\sim 0 \nonumber
\end{align}  
where $ n^{(1)\beta \beta}_{L} $ is given by \eqref{eq:N1bb}.  Again, these prefactors can be understood by simply distinguishing local shear versus convergence information, together with parity arguments. The factors are the same as for the estimator responses. Hence, the relative importance of the $N^{(1)}$ to $N^{(0)}$ biases at low-$L$ is exactly the same for the three $\beta$, $\omega$ and $\kappa$ estimators.

In the delensed case, where most of the lensing power has been removed, the squeezed analytic approximation is not expected to hold, and we use the same flat-sky expressions but with the partially delensed spectra and delensed $C_L^{\psi\psi}$ as input. These are shown in orange on Fig.~\ref{fig:N1comp}. As can be expected, the relevance of the $N^{(1)}$ term is significantly reduced compared to the QE case. Now the $N^{(1)}$ bias is more than an order of magnitude smaller than $N^{(0)}$. This shows the power of delensing. Hence we neglect the $N^{(1)}$ biases from now on.

\begin{figure}
\centerline{
\includegraphics[width=0.7\linewidth]{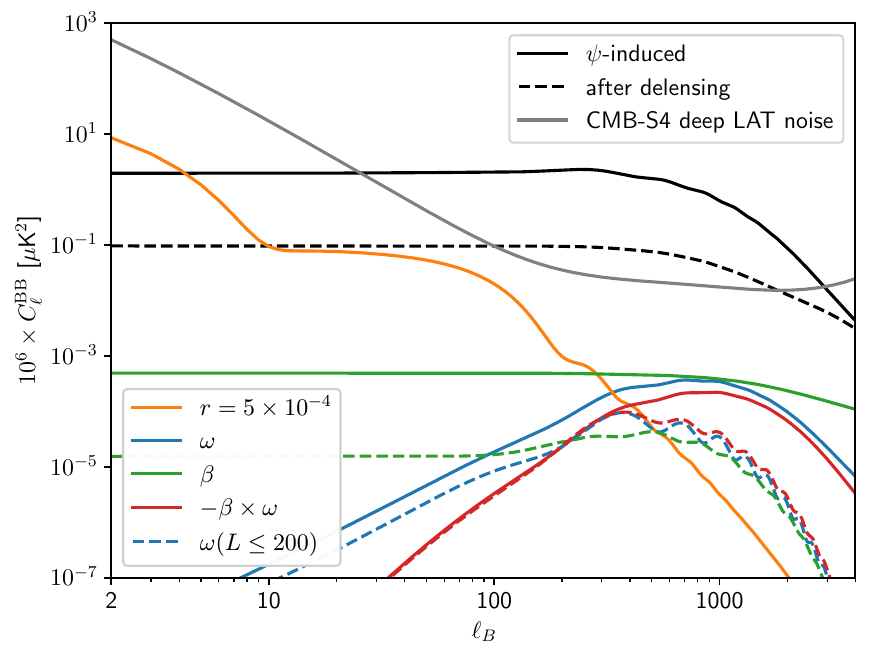}
 	}
	\caption{\label{fig:BBcomp}
	Comparison of various $B$-mode power spectra. Black solid is the standard, gradient lensing contribution, and dashed black after delensing. (Delensing is shown here for CMB-S4 deep, with residual effective noise and foreground spectrum shown by the grey line). Delensing is performed iteratively as described in the main text, achieving a removal of  $95\%$ to $85\%$ on the scales relevant for the reconstructions of the lensing field rotation and polarization rotation.
	The blue line shows the contribution from lensing field rotation ($\omega$, blue), and the green line the effect of a polarization rotation ($\beta$), in the model where its spectrum is equal to that of $\omega$. If the polarization rotation is equal to $-\omega$, the power gets an additional contribution shown as the red line. These are the contributions to B-polarization from lensing rotation as obtained in Ref.~\cite{DiDio:2019rfy}. Neither the lensing contribution $\om$ not the cross term $\om\times\beta$ have a significant impact on the quality of the reconstructions, which take most of the signal from a broad kernel $\sim 500 \leq \ell \leq 2000$, where the effective number of modes is highest. (There are bispectrum-sourced terms at the same order, but these are even smaller on the scales relevant here and are not shown.) The dashed colored lines show the contributions to the B-modes of the large scale rotation modes exclusively $(L \leq 200)$, as relevant for the reconstruction on these scales.
 }
\end{figure}

A high-resolution $B$-mode map is key for best reconstruction performance. For orientation, we show in Fig.~\ref{fig:BBcomp} various contributions to the $B$-mode power. The standard $\La$CDM gradient lensing term of $\sim 5\mu \text{K}$arcmin is shown as the black line. The grey line shows the residual effective noise according to our modelling of the CMB-S4 deep patch Large Aperture Telescope (LAT). This is obtained with an internal linear harmonic combination of the six frequency channels, using a parametric synchrotron and polarized dust foreground model that follows the \BK implementation and
constraints~\cite{2022ApJ...926...54A,BICEP2:2018kqh}. Our experimental configurations are discussed in more detail in Appendix~\ref{ap:exp}.

The dashed black line shows the residual lensing power achievable after delensing in this configuration.
The amount of delensing is scale-dependent, with residual lensing power approximately $5\%$ for $\ell < 500$, increasing to $15\%$ at $\ell = 2000$ (where LAT noise starts to dominate) and $70\%$ at $4000$ respectively.

The blue line shows the $B$-mode power induced by the lensing rotation field $\omega$. The green line is that induced from polarization rotation, assuming $\beta$ is a Gaussian field with spectrum identical to $C_L^{\omega\omega}$. If $\beta$ is in fact equal to $ -\omega$, there is also a cross-term induced by $C_L^{\beta \omega}$. This contribution is shown as the red line, boosting slightly the power around the peak (all these curves were obtained using the full-sky correlation function approach~\cite{Challinor:2005jy, Lewis:2006fu}). They are in good agreement with previous results~\cite{DiDio:2019rfy}. The total sum of these three contributions remains well below the residual effective noise + delensed power, and we have neglected all of them when building quadratic estimator reconstruction noise spectra. The B-modes produced by the polarization rotation $\beta$ and the lensing curl $\omega$ are somewhat correlated on all scales, but this is mostly true for those produced by the largest scale rotation modes. The colored dashed lines show the contribution coming from  $L\leq 200$; these B-modes have a cross-correlation coefficient of -0.8 inducing a significant degenaracy between $\beta$ and $\omega$ reconstruction on these scales.

\section{Prospects for detection}\label{s:det}

\subsection{CMB internal spectrum reconstruction}\label{ss:recnoise}
\begin{table}
\centering
\begin{tabular}{|l|c||c|c|c||c||c|}
\hline
     &$100\cdot f_{\text{sky}}$ & $\beta$ & $\omega $& $\beta = -\omega$ & $\kappa$  & $\sigma_{\bar \beta} [\textrm{arcmin}]$  \\
   \hline
   LiteBird   & 60  &     &     &     & 63  & 0.44  \\
  SO-baseline (P-only) & 40  &     &     &     & 82  & 0.51 \\
  SO-goal (P-only)    & 40  & 0.1 & 0.1 & 0.1 & 117 & 0.35 \\
  SPT-3G-7y & 3.6 & 0.2 & 0.2 & 0.2 & 91 & 0.38   \\
  PICO & 60 & 1.4 & 1.6 & 1.7 & 475 & 0.06\\
  S4-wide     & 40  & 1.3 & 1.0 & 1.5 & 379 & 0.08  \\
  S4-deep     & 3.6   & 3.5 & 2.6 & 3.9 & 232  & 0.1\\
  \hline
\end{tabular}\caption{SNR on internal auto-spectrum detection, computed from \eqref{eq:SNR}, and using polarization data only. All numbers include internal iterative delensing self-consistently. Columns 3 and 4 treat $\om$ and $\beta$ independently (and having the same power spectrum) while in column 5 we set $\beta=-\om$. In column 6 we show the gradient-mode lensing SNR for comparison (which is signal dominated on a wide range of scales, hence the S4-wide survey is most efficient) and in the last column we present the $1\si$ error on a constant polarization rotation angle $\beta$. }
\label{table:SNRsauto}
	\end{table}

Next we discuss prospects for purely internal reconstruction.
We compute signal to noise ratios for power spectrum reconstruction, assuming Gaussian statistics,\begin{equation}\label{eq:SNR}
	(S/N)^2 = \frac 12 f_{\rm sky} \sum_{L \ge 30} (2L + 1)\left( \frac{C^{\omega \omega}_L}{C_L^{\omega \omega} + N_L^{(0)}} \right)^2.
\end{equation}

The numbers are given in Table~\ref{table:SNRsauto}, for different experiments.
More details on the configurations  are found in Appendix~\ref{ap:exp}. All results are for polarization only and use iterative internal delensing as described in the previous section, making the deep configuration much better for both angles.
The first two columns after the vertical double line ($\beta$ and $\omega$) treat the two fields as independent but assume the same signal spectral shape $C_L^{\omega\omega}$, and the third assumes $\beta = - \omega$.
In this case, for the deep configuration, there is less information for $L \leq 700$, but the signal is boosted above that, increasing overall slightly the expected significance since there are more modes there. 
In this deep configuration such a signal is potentially detectable at $3.9 \sigma$. A similar result has found by an order-of-magnitude estimate presented in~\cite{Fabbian:2017wfp}. In the last column we also show the $1\si$ error on a constant polarization rotation. For this S4-wide and especially PICO are best due to their large sky coverage which helps in detecting the monopole of $\beta$.

\subsection{External rotation templates}\label{s:templates}
Since the lensing curl arises from multiple (nonaligned) lenses, we can leverage cross-correlation with large-scale structure tracers, which provide a better redshift resolution and reduced noise relative to the CMB auto-correlation. Considering an external template for lensing rotation $\omega$, we can significantly improve the signal-to-noise ratio.
Before investigating the optimal choice of LSS tracers to build a rotation template, we determine the signal-to-noise ratio in the ideal case of having a perfect lensing rotation template. Assuming Gaussian statistics, the signal-to-noise ratio is given by
\begin{equation} \label{eq:SNR_cross}
    	(S/N)^2 = f_{\rm sky} \sum_{L \ge 30}^{L_{\rm max}} (2L + 1) \frac{C^{\omega \omega}_L}{2 C^{\omega \omega}_L + N_L^{(0)}} \, .
\end{equation}
\begin{table}[ht]
\centering
\renewcommand{\arraystretch}{1.2} % Adjust this value to reduce/increase row separation

\begin{tabular}{|l|c||c|c|c||}
\hline
     &$100\cdot f_{\text{sky}}$ & $\omega$ & $\beta $& $\beta = -\omega$  \\
   \hline
   LiteBird   & 60  & 2.1    &  2.9   &   3.3    \\
  SO-baseline (P-only) & 40  &  6.1   &  11.5   &  12.3    \\
    SO-baseline (GMV) & 40  &  9.9   &  11.4   &  14.5    \\
  SO-goal (P-only)    & 40  & 9.1 & 16.8 & 18.1  \\
    SO-goal (GMV)    & 40  & 12.1 & 16.8 & 19.7  \\
  SPT-3G-7y & 3.6 & 8.0 & 14.0 & 14.9    \\
  PICO & 60 & 44.8 & 62.6 & 68.5 \\
  S4-wide     & 40  & 37.0 & 64.2 & 68.5   \\
  S4-deep     & 3.6   & 33.4 & 58.9 & 62.3 \\
  \hline
\end{tabular}
\caption{
The S/N ratio is shown for an ideal exact lensing rotation template.
In the third column,  `$\omega$', we consider only image rotation  as done in Ref.~\cite{Robertson:2023xkg}. In the fourth column,`$\beta$', we consider only polarization rotation assuming both, $\om$ and $\beta$ have the same spectrum. Finally in the fifth column, `$\beta=-\omega$',  we consider both and set $\beta=-\om$ as advocated in Refs.~\cite{Marozzi:2016qxl,Marozzi:2016und,DiDio:2019rfy}.}
\label{table:SNRcross}
\end{table}

\begin{figure}
	\includegraphics[width=\columnwidth]{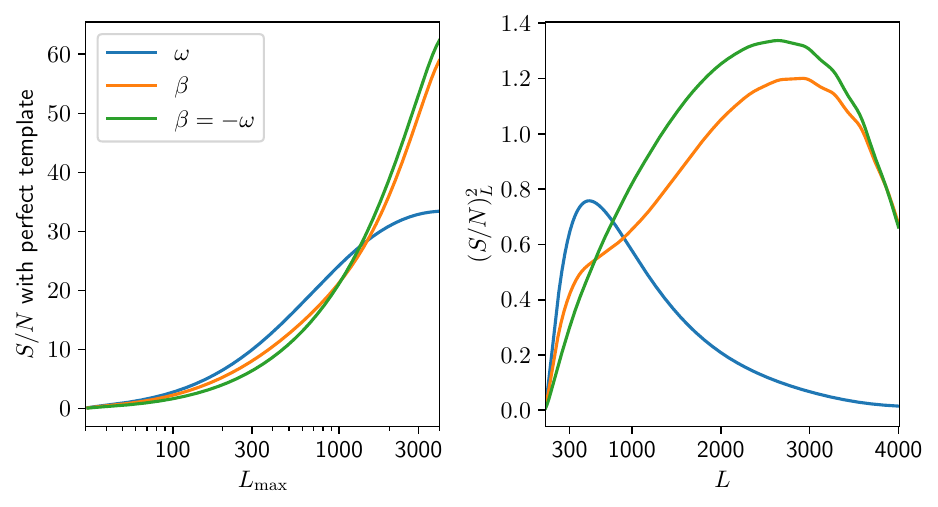} 
 \caption{We show the cumulative S/N ratio (left) and the S/N square per mode (right) for CMB S4-deep ($f_{\rm sky}=0.036$) in an ideal case, i.e.~with a perfect $\omega$ template, as a function of $L_{\rm max}$ for image rotation (blue), polarization rotation (green) and both of them with $\beta  =- \omega$ (red).}
 \label{fig:SN_cumulative} 
 \end{figure}

From Table~\ref{table:SNRcross} we see that having an exact ideal lensing rotation template would lead to a detection already with the SO baseline, and even earlier with SPT-3G, which is taking data right now. Taking into account the polarization rotation with $\beta = - \omega$ the S/N ratio was roughly doubled for any configuration considered here. In this ideal case, all configurations except LiteBird would detect such a signal with $S/N>12$. However, such an improvement requires us to reach scales much smaller than $L \sim1000$. Indeed, as we see in Fig.~\ref{fig:SN_cumulative}, the presence of the polarization rotation with $\beta = -\omega$ reduces $S/N$ somewhat to scales $L \lesssim 1000$. This is due to the negative contribution of $C_\ell^{\om\beta}$ to $C_\ell^{BB}$.
Looking at the signal-to-noise ratio per mode, we see that with  $\omega$ only, it peaks around $L \sim 500$, while including also the polarization rotation $\beta$ it peaks at much smaller scales around $L\sim 2000$.

\subsection{Templates for the lensing curl from LSS}\label{s:lss}
We have seen that, with a perfect rotation template the lensing rotation can be detected with upcoming surveys. In this section we aim to build a rotation template from LSS tracers. The goal is to obtain a template which is as much correlated as possible with the true lensing rotation. Indeed in the presence of a non-perfect template the signal to noise is degraded to 
\begin{equation}
    (S/N)^2 = f_{\rm sky} \sum_{L \ge 30}^{L_{\rm max}}  \left( 2 L+ 1\right) \frac{C^{\omega \omega}_L}{\left( C^{\omega \omega}_L + N^{(0)}_L \right) F^{-1}_L + C^{\omega \omega}_L }
\end{equation}
where $\sqrt{F_L}$ denotes the correlation coefficient between the template and the true rotation field. With $F_L =1 $ we recover the ideal signal-to-noise ratio eq.~\eqref{eq:SNR_cross}. 
The generic value of $F_L$ is introduced below.

Following Ref.~\cite{Robertson:2023xkg}, we  determine a rotation estimator template $\hat \omega$, normalized such that it has minimal variance and is unbiased.
We introduce a general weighted quadratic estimator\footnote{We use flat-sky approximation only to derive the correlation coefficient $F_L$. As shown in Ref.~\cite{Robertson:2024qjl}, this approximation provides a reliable comparison with full-sky numerical simulations.}
\begin{equation}\label{e:36}
    \hat \omega \left( \bL \right) = \frac{F_L^{-1}}{2} \int \frac{d^2L_1 }{2 \pi } G_{ij} \left( \bL, \bL_1 \right) \hat a^i \left( \bL_1 \right) \hat a^{j}\left( \bL - \bL_1 \right)
\end{equation}
for a set of LSS tracers $\left\{ \hat a^i \right\}$. The weights $G_{ij} \left( \bL, \bL_1 \right)$ and the normalization constant $F_L$ are set such that
\begin{eqnarray} \label{eq:cond1}
\langle \hat \omega \left( \bL \right) \omega \left( \bL_1 \right) \rangle &=& \delta_D \left( \bL + \bL_1 \right) C^{\omega \omega}_L  \, ,\\
\langle \hat \omega \left( \bL \right) \hat \omega \left( \bL_1 \right) \rangle &=& \delta_D \left( \bL + \bL_1 \right) C^{\omega \omega}_L F_L^{-1} \, .
 \label{eq:cond2}
\end{eqnarray}
The first condition guarantees that $\hat\om$ is unbiased, while the second leads to minimal variance~\cite{Robertson:2023xkg}.
Eq.~\eqref{eq:cond1} implies
\begin{equation} \label{eq:FL_eq1}
    F_L = \frac{1}{2 C^{\omega \omega}_L} \int \frac{d^2L'}{\left( 2 \pi \right)^2} G_{i j} \left( \bL, \bL' \right) b^{\omega i j}_{-\bL, \bL', \bL - \bL'}\, ,
\end{equation}
where we have introduced the bispectrum between the lensing curl $\omega$ and the LSS tracers
\begin{equation}
  \langle   \omega \left( \bL \right) a^{i} \left( \bL_1\right) a^j \left(\bL_2 \right) \rangle = \frac{1}{2\pi}
 \delta_D\! \left( \bL\! +\! \bL_1\! +\! \bL_2 \right) b^{\omega i j}_{\bL \bL_1 \bL_2} \, .
\end{equation}
Eq.~\eqref{eq:cond2} leads to
\begin{align} \label{eq:FL_eq2}
 &  C_L^{\omega \omega} F_L = \frac{1}{4} \int \frac{d^2L'}{\left( 2\pi \right)^2} G_{ij} \left( \bL, \bL'\right)
 \left[
   G_{pq} \left( - \bL , - \bL' \right) C^{ip}_{L'} C^{jq}_{\left| \bL - \bL' \right|}
   +
   G_{pq} \left( - \bL, \bL' - \bL \right) C^{iq}_{L'} C^{jp}_{\left| \bL - \bL' \right|}
    \right] \, ,
\end{align}
where $C^{ij}_{L}$ denote
 the spectra of the LSS tracers,
\be\label{e:CLSS}
\langle a^{i} \left( \bL\right) a^j \left(\bL' \right) \rangle =
 \delta_D \left(\bL + \bL' \right)C^{ij}_{L} \,.
\ee
Combining equations~(\ref{eq:FL_eq1}) and \eqref{eq:FL_eq2} we obtain
\begin{equation}
    b^{\omega i j}_{-\bL, \bL', \bL - \bL'} = G_{pq} \left( - \bL, - \bL' \right) C^{ip}_{L'} C^{jq}_{\left| \bL- \bL' \right|}
\end{equation}
and
\begin{equation}
    F_L = \frac{1}{2C^{\omega \omega}_L} \int \frac{d^2L_1}{\left( 2 \pi \right)^2} b^{\omega i j}_{-\bL, \bL_1, \bL_2} b^{\omega p q}_{-\bL, \bL_1, \bL_2} \left( C^{-1}_{\rm LSS} \right)^{ip}_{L_1} \left( C^{-1}_{\rm LSS} \right)^{jq}_{L_2} \ ,
\end{equation}
with $\bL_2 = \bL - \bL_1 $.
The curl estimator in terms of the bispectrum is then given by
\begin{equation}\label{eq:curl_est_bisp}
    \hat \omega \left( \bL \right) = \frac{F_L^{-1}}{2} \int \frac{d^2L_1}{2 \pi } b^{\omega i j}_{-\bL, \bL_1, \bL_2}
\left( \mathbf{C}^{-1}_{\rm LSS} \hat {\mathbf{a}} \right)^i_{\bL_1 }  \left( \mathbf{C}^{-1}_{\rm LSS} \hat {\mathbf{a}} \right)^j_{\bL_2 } \, .
\end{equation}
where in the inverse covariance we include also the noise $N^{ij}_L$ of the LSS tracers:
\begin{equation}
   \left(  \mathbf{C}_{\rm LSS}\right)^{ij}_L= C^{ij}_L + N^{ij}_L \, .
\end{equation}

In Appendix~\ref{app:bisp}, we compute the bispectrum between the lensing curl and the LSS probes, finding
\begin{align}
&   b_{(-\bL) \bL_1 \bL_2}^{\omega i j}
         =  2  \frac{\bL_1 \times \bL_2}{L_1^2 L_2^2} 
         \left( \bL_1 \cdot \bL_2 \right)
 \left[ M^{i j} \left( L_1,L_2 \right) -   M^{j i} \left( L_2,L_1 \right)\right] \, . \nonumber
\end{align}
where the the coefficients $M^{i j} \left( L_1,L_2 \right)$ are defined in eq.~\eqref{eq:M_coeff_bisp}.

In our analysis we consider 3 different LSS tracers: the CMB lensing convergence $\kappa$, galaxy clustering $g$ and the cosmic infrared background $I$. These tracers $\left\{a^i \right\}$ are related to the density fluctuation $\delta$ through the line-of-sight integral
\begin{equation} \nonumber
    a^i \left( \bL \right) = \int dr W_i \left( r \right) \delta \left( \bL, r \right)
\end{equation}
where the window functions $W^i$ are given by
\begin{eqnarray}
\label{eq:Wkappa}
W_\kappa (r) &=& \frac{3 \Omega_M H_0^2}{2 a(r)} r^2 \frac{r_*- r}{r r_*} \, ,\\
\label{eq:Wg}
    W_g (r) &=& b_1 (r) n(r) H(r) \, ,\\
    \label{eq:WI}
     W_I (r) &=& A_\nu \frac{r^2}{\left( 1 + z(r) \right)^2} \exp \left( - \frac{\left( z - z_c \right)^2}{2 \sigma_z^2}\right) f_{\nu (1+ z)} \, .
\end{eqnarray}
More details about these LSS tracers which agree with the ones used in~\cite{Robertson:2023xkg} can be found in Appendix~\ref{app:LSS}.

\begin{figure}
	\includegraphics[width=0.49\columnwidth]{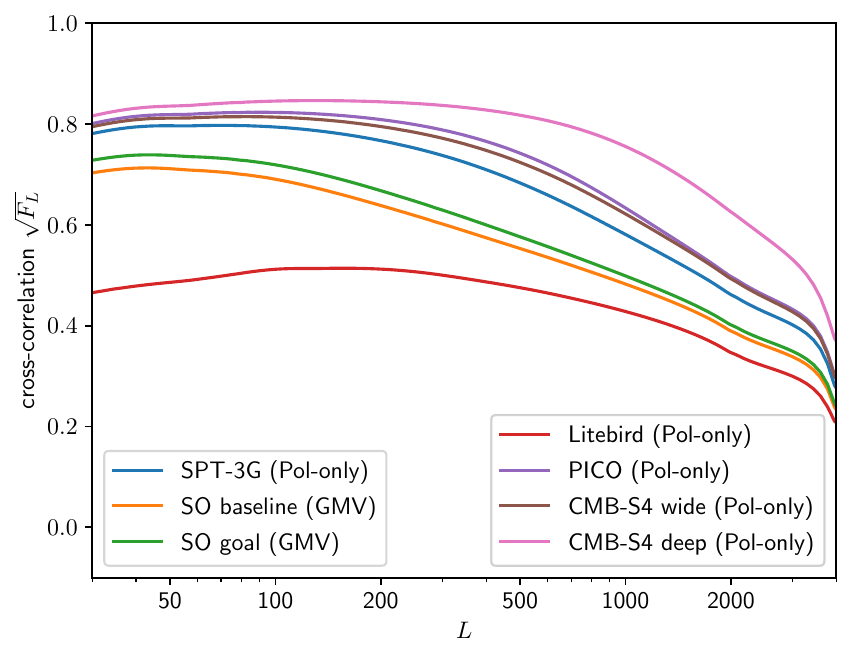}
\includegraphics[width=0.49\columnwidth]{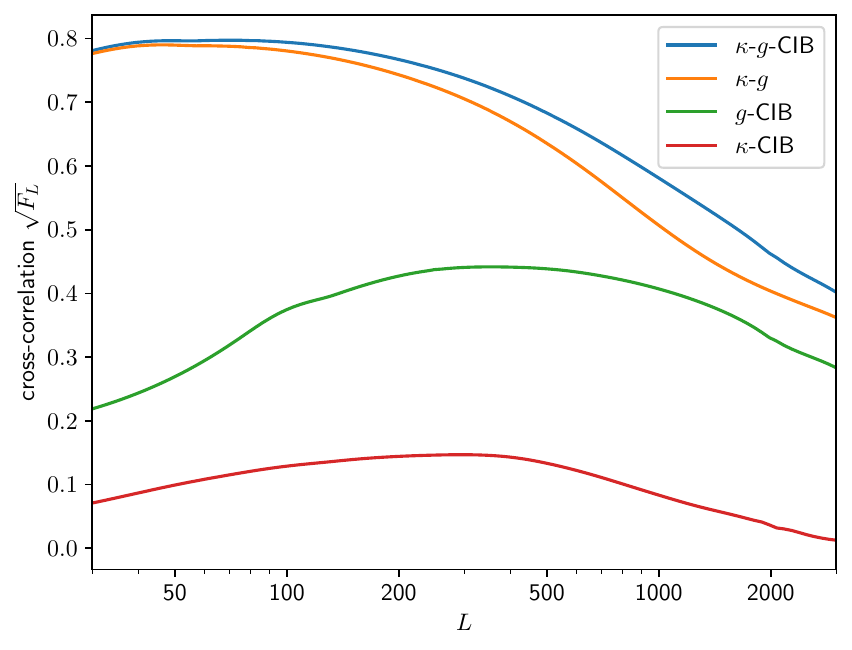}

\caption{\emph{Left panel:} Cross-correlation coefficients \(\sqrt{F_L}\) of curl lensing templates from LSS are plotted for different CMB experiments. The LSS templates are constructed using CMB lensing convergence, galaxy clustering, and CIB. CMB-S4 deep shows the highest correlation with a perfect curl template, reaching approximately 0.8 at large scales and maintaining high correlation on small scales longer than any other survey. \emph{Right panel:} The different contributions to the correlation coefficient \(\sqrt{F_L}\) for the curl lensing template from the same LSS tracers  are plotted for SPT-3G-7y. Most of the correlation is driven by the combination of lensing convergence and galaxy clustering, with CIB contributing only marginally.}
 \label{plot:FL}
 \end{figure}

Evaluating the bispectrum between the lensing curl and the LSS tracers, we obtain the correlation coefficient $F_L$. The square root of $F_L$ determines the correlation coefficients of the different LSS templates with a perfect lensing rotation template. In fig.~\ref{plot:FL} we show the correlation coefficients $\sqrt{F_L}$ as a function of the multipoles $L$ for different LSS tracers.  Using these correlation coefficients, we have computed the signal to noise ratio to detect the lensing curl through LSS cross-correlations. The results are shown in tables~\ref{table:SNRcross_omega}-\ref{table:SNRcross_beta_omega}.
In table~\ref{table:SNRcross_omega}, we assume  that the image is rotated, but not the polarization. Under this assumption, our results are consistent with Ref.~\cite{Robertson:2023xkg} when using the same survey specifications. In this case we find that an initial detection (SNR$\sim5$) is forecasted with SPT-3G by optimally combining a LSS template with lensing convergence $\kappa$, galaxy clustering, and CIB.
With PICO or S4 (wide and deep), lensing rotation can be detected at high significance whenever a galaxy clustering template for $\om$ is used. $S/N$ is between 12 and 24 depending on the LSS tracer used.

Finally in table~\ref{table:SNRcross_beta_omega} we consider that both the image and the polarization are rotated with $\beta = -\omega$. For this case, we will reach SNR$\;\sim 7$ already with SPT-3G, while the next generation of surveys (PICO and S4) will detect the lensing curl with a very good accuracy.

\begin{table}[ht]
\centering
\renewcommand{\arraystretch}{1.2} % Adjust this value to reduce/increase row separation

\begin{tabular}{|l|c||c|c|c|c||}
\hline
     &$100\cdot f_{\text{sky}}$ & $\omega [\kappa,g]$ & $\omega  [\kappa,I]$ & $\omega [g,I]$ & $\omega [\kappa,g,I]$  \\
   \hline
   LiteBird  & 60  &  0.7  &  0.1   &  0.9 &  1.1   \\
  SO-baseline (P-only) & 40  &  1.8  &  0.4 & 2.5 & 2.9  \\
    SO-baseline (GMV) & 40  & 3.4  & 0.6  & 3.7 & 4.6   \\
  SO-goal (P-only)    & 40  &  3.3  & 0.6  &  3.6& 4.5\\
    SO-goal (GMV)     & 40  &  4.7  & 0.8  & 4.6 & 6.0 \\
  SPT-3G-7y & 3.6 & 4.5   &0.8 &  3.2  &  4.9 \\
  PICO & 60 &  29.5  & 5.6  & 18.6 & 31.2 \\
  S4-wide     & 40 & 22.4   & 4.1 & 14.8 & 23.9   \\
  S4-deep     & 3.6   &  24.0  & 4.2 & 13.3 & 24.9 \\
  \hline
\end{tabular}
\caption{We show the S/N ratio for the cross-correlation between CMB reconstructed field rotation $\hat \omega$ and different LSS rotation templates $\hat \omega^{\rm template}$. Here we assume only the image but not the polarization is rotated. This corresponds to column 3 of Table~\ref{table:SNRcross} but with realistic lensing rotation templates.}
\label{table:SNRcross_omega}
\end{table}
\begin{table}[ht]
\centering
\renewcommand{\arraystretch}{1.2} % Adjust this value to reduce/increase row separation
\begin{tabular}{|l|c||c|c|c|c||}
\hline
     &$100\cdot f_{\text{sky}}$ & $\beta = - \omega[\kappa,g]$ & $\beta = - \omega [\kappa,I]$ & $\beta = - \omega [g,I]$ & $\beta = - \omega [\kappa,g,I]$  \\
   \hline   LiteBird   & 60  &   0.9 &0.2&  1.4 &   1.6  \\
  SO-baseline (P-only) & 40  &  2.6  & 0.4  & 4.1 & 4.6  \\
    SO-baseline (GMV) & 40  &  4.3 & 0.6   & 4.9 & 6.0  \\
  SO-goal (P-only))    & 40  &  4.7  &  0.7 & 6.0 & 7.0 \\
    SO-goal (GMV)    & 40  & 6.3   & 0.9  &  6.6& 8.3 \\
  SPT-3G-7y & 3.6 & 6.4   & 0.8& 5.0  & 7.1  \\
  PICO & 60 &    36.6&  5.7 & 25.0  &39.4\\
  S4-wide     & 40 & 32.0    & 4.1&  22.7 & 34.7  \\
  S4-deep     & 3.6   & 37.3  & 5.1 & 20.4  & 38.3 \\
  \hline
\end{tabular}
\caption{We show the S/N ratio for the cross-correlation between CMB reconstructed field rotation $\hat \omega$ and $\hat \beta$  and different LSS rotation templates $\hat \omega^{\rm template}$. Here we consider both image rotation and polarization rotation with $\beta=-\om$ in the CMB. This corresponds to column 5 of Table~\ref{table:SNRcross} but with realistic lensing rotation templates.}
\label{table:SNRcross_beta_omega}
\end{table}

\begin{figure}
\centering
	\includegraphics[width=0.89\columnwidth]{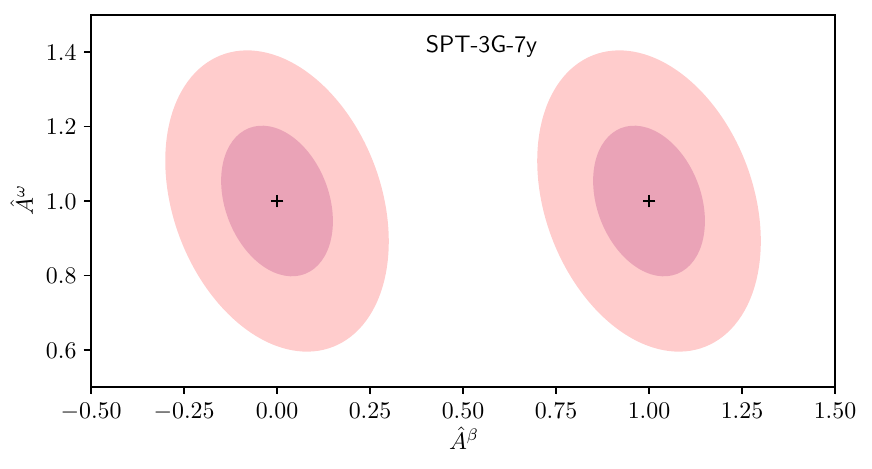}
\caption{Forecast $1\sigma$ and $2\sigma$ contours on a joint measurement of the amplitude of the lensing curl mode power and of the polarization rotation power, under two different scenarios. On the right, the Universe has $\beta = -\omega$, as predicted by Refs~\cite{Marozzi:2016und,Marozzi:2016qxl,DiDio:2019rfy}, and both signals are clearly jointly detected. If $\beta$ is much smaller, as predicted by Refs~\cite{Lewis:2016tuj,Lewis:2017ans,Namikawa:2021obu} (effectively undetectable), then the expected output of the same measurement procedure is the ellipse on the left. In both cases this considers the cross-correlations of the $\beta$ and $\omega$ estimates from SPT-3G CMB to the best templates considered in this work (`$\kappa, g, I$'). The degeneracy (here a cross-correlation coefficient of $\approx -0.27$ for both ellipses) is set by the overlap of the $\beta$ and $\omega$ reconstruction noise at small and moderate $L$ (see Fig.~\ref{fig:N0comp}, right panel). The ellipses are essentially identical, since the different $\beta$ signal has only a tiny effect on the variances and covariance, all dominated by Gaussian reconstruction noise.}
 \label{fig:ellipse}
 \end{figure}

\section{Conclusion}\label{s:con}

Very precise CMB polarization measurements allow excellent lensing reconstruction as well as delensing. This makes very deep experiments rather than wide area surveys in principle the best places to look for small signals limited by lensing sampling variance. 

We have looked in some details on the prospects for the measurement of lensing field rotation as well as polarization rotation, which both up to now usually serve as diagnostic of systematics. Lensing and polarization rotation quadratic estimators behave very similarly in this regime, and we have discussed in details their reconstruction noise and degenaracies (on large-scales purely rotating (`non-shearing') lenses cannot be told apart from a polarization rotation).
 The PICO and CMB-S4 experiments can constrain a constant rotation angle to about  20 arcseconds at $5\si$.
This small number is not quite enough to extract from the CMB alone the guaranteed post-born lensing field rotation signal with great significance, in agreement with previous work~\cite{Robertson:2023xkg}. 
If the polarization rotation angle induced by large-scale structure is similar to that of lensing rotation, as recent works claim~\cite{Marozzi:2016uob,Marozzi:2016und,DiDio:2019rfy}, then the reconstruction noise is larger on the largest scales due to the partial cancellation of $B$-modes produced by large lenses and large-scale polarization rotations. However, the total SNR on a detection of the spectrum is boosted owing to the more efficient production of $B$-modes by small scale polarization rotation. Internal detection remains challenging, with a forecast detection of about $3.9\sigma$ when including state-of-the-art delensing. In practice, this `delensing' can be achieved for example by joint Maximum A Posteriori reconstruction of $\kappa$, $\omega$ and/or $\beta$, which automatically removes the $\kappa$-induced variance from the rotation estimates. (See \cite{Belkner:2023duz} for joint $\kappa$-$\omega$ reconstructions which achieve expectations. The case of $\beta$ is similar and will be presented elsewhere).

On the other hand, cross-correlating of the rotation signal with templates for the field rotation $\om$ from large scale structure observations
lead to a detection with $S/N$ up to 39 for PICO and 38 for CMB-S4 deep. Interestingly, already SPT-3G-7y has   $S/N=7$ for the case of a rotation of both, the lensing image and polarization which is very promising.

Even if workers in the field cannot agree whether image rotation equivalently generates a polarization rotation or not, near future data can tell. See Fig.~\ref{fig:ellipse}.

Polarization rotation can also be induced by other physical effects than lensing. E.g. by Faraday rotation in the presence of an external magnetic field~\cite{Kosowsky:2004zh} or by coupling to a spatially constant or varying axion field.

The B-mode polarization spectrum induced by a magnetic field peaks at much smaller scales, namely $\ell\simeq 13000$, see~\cite{Kosowsky:2004zh}, than the one induced by lensing rotation. Furthermore, as the Faraday rotation angle is frequency dependent, behaving as $1/\nu^2$, a detailed study of limits on magnetic fields from this effect for different CMB experiments, will also depend on the frequencies observed with the experiments under consideration and not only on the magnetic field spectrum.

Constraints on a scale invariant rotation angle spectrum by future CMB experiments have been derived in~\cite{Pogosian:2019jbt,Zhong:2024tgw}, and~\cite{Pogosian:2019jbt} also quotes limits on a constant polarization rotation angle.

Finally, it is interesting to note that a  constant polarization rotation
of $\beta_0= (0.342\pm 0.094)^o$ is tentatively detected in the Planck and WMAP data, see~\cite{Eskilt:2022cff} and refs. therein. Provided the very difficult systematics can be managed, this can be confirmed (or rejected) at extremely high significance as the error on $\beta_0$ will be of the order of $\De\beta_0\sim 4.8'' \simeq 0.0013^o$ for the CMB-S4 and even $\De\beta_0\sim 3.6'' \simeq 0.001^o$ for PICO.

\begin{acknowledgments}
We thank M. Robertson, A. Lewis and G. Fabbian for discussions at various stages of this work. We acknowledge support from the Swiss National Science Foundation. We thank the anonymous referee for a number of useful comments and suggestions.
JC acknowledges support from a SNSF Eccellenza Professorial Fellowship (No. 186879).
ED acknowledges funding from the European Research Council (ERC) under the European Union Horizon 2020 research and innovation program (Grant agreement No.~863929; project title “Testing the law of gravity with novel large-scale structure observables).
\end{acknowledgments}
\appendix

\section{The CMB experiments considered in this work}\label{ap:exp}
\begin{figure}
	\centerline{\includegraphics[width=0.7\columnwidth]{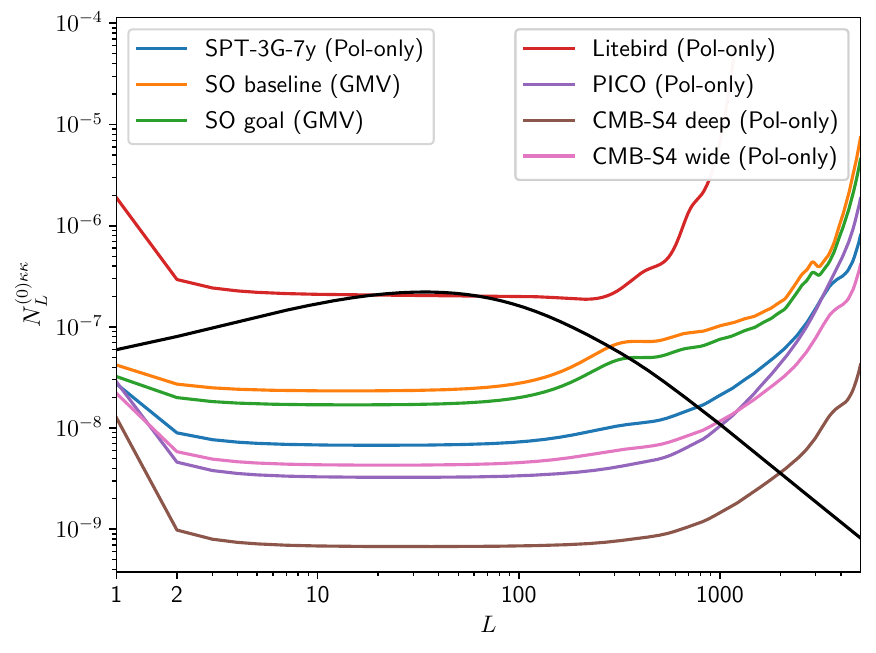}} 
 \caption{Lensing convergence reconstruction noise levels $N_L^{(0)}$ for the CMB configurations used in this work. The black line is the fiducial lensing power spectrum. Curves for Simons Observatory combine temperature and polarization information, the others use polarization only.}
 \label{fig:N0kks} 
 \end{figure}
The Simons Observatory (SO) noise curves are the official baseline and goal noise curves taken from\footnote{\url{https://github.com/simonsobs/so_noise_models}}~\cite{SimonsObservatory:2018koc}. For the other experiments, we use tables of white noise levels, and of $\ell_{\rm knee}$ and exponents for the non-white part for each frequency from public sources. The numbers are taken from \cite{SPT-3G:2024qkd} (SPT-3G main survey), \cite{Alvarez:2019rhd}(PICO), \cite{LiteBIRD:2022cnt}(LiteBird), \cite{Belkner:2023duz}(CMB-S4 deep), \cite{CMB-S4:2020lpa}(CMB-S4 wide). For the deepest configuration in polarization we further include at each frequency galactic polarized synchrotron and galactic dust foreground emission, using for this the same parametric model and fiducial values detailed in the appendix of~\cite{CMB-S4:2020lpa} which are based on \BK observations~\cite{BICEP2:2018kqh}. This does not play an essential role however since we focus on small enough scales. In temperature, we use multipoles below $\ell=3000$, in which case the impact of foreground can probably be mitigated. We combine the frequency channels using a minimum variance internal linear combination to produce a single noise curve from which we compute reconstruction noise levels with and without delensing etc.

\section{LSS tracers \label{app:LSS}}
In this appendix, we summarize the specifications of the LSS tracers used in the analysis. To facilitate a direct comparison with the results of Ref.~\cite{Robertson:2023xkg}, we adopt the same choice of LSS tracers.

In the S/N ratio we need to evaluate the angular power spectrum of the LSS tracers. In terms of generic LSS tracer window functions $W_{i}$, using the Limber approximation (see e.g.~\cite{Durrer:2020fza}) the angular power spectrum is given by
\begin{equation}
    C_L^{ij}= \int  dr \frac{W_{i} (r)W_{j} (r)}{r^2} P_{\delta \delta} \left( \frac{L}{r}, z(r) \right)\, ,
\end{equation}
where we adopt Halofit~\cite{Takahashi:2012em} for the matter power spectrum. In computing the bispectrum with the curl lensing potential, we also need to evaluate the angular power spectrum between the lensing convergence $\kappa$ and the LSS tracers at different comoving distance $r$
\begin{equation}
    C_{L}^{\kappa i} = \frac{3 \Omega_M H_0^2}{2}\int_0^r dr' \frac{r - r'}{r r'} \frac{ W_{i} (r') }{a(r')}P_{\delta \delta} \left( \frac{L}{r'}, z(r') \right)\, .
\end{equation}
For this we use $\ka=-\frac 12\De\psi$ and the Poisson equation relating $-k^2\Phi = 4\pi G\rho_ma^2\de$. 

\subsection{CMB lensing convergence}
For CMB lensing convergence we consider different upcoming surveys, summarized in tables~(\ref{table:SNRsauto}-\ref{table:SNRcross_beta_omega}). In forecasting the signal to noise to detect $\omega$ and/or $\beta$ we consider the same CMB experiment for the lensing curl rotation and $\kappa$ as a LSS  probe. Beyond the window function, see eq.~\eqref{eq:Wkappa}, we simply need the noise curves $N^{(0)}_L$, shown in fig.~\ref{fig:N0kks}.

\subsection{Galaxy Clustering}
The second LSS probe we consider is galaxy clustering. Given that the lensing kernel is quite broad, we do not gain significant benefits from using spectroscopic surveys. Instead, we can take advantage of the very low shot noise provided by photometric surveys such as the Vera Rubin Observatory (formerly LSST)~\cite{LSSTScience:2009jmu}. For a broad redshift bin we can neglect redshift space distortions and describe galaxy clustering with help of the window function defined in eq.~\eqref{eq:Wg}. The linear bias parameter is modelled as
\begin{equation}
    b(z) = 1+ 0.84 z \, ,
\end{equation}
while the galaxy redshift distribution is
\begin{equation}
    \frac{dN}{dz} = \frac{1}{2 z_0} \left( \frac{z}{z_0} \right)^2 e^{- z/z_0}
\end{equation}
with $z_0=0.311$ and $n(z)$ is the corresponding normalized (to unity) distribution.
The noise is dominated by shot-noise, determined by the number of galaxy per steradian
\begin{equation}
    N_{\rm shot} =  \frac{1}{40} \left(\frac{\pi}{60 \times 180} \right)^2 \simeq 2.1 \times 10^{-9} \, .
\end{equation}

\subsection{Cosmic Infrared Background}
  For CIB radiation we consider the signal spectral energy distribution model, see Ref.~\cite{Hall:2009rv}. The function $f_\nu$ which characterize the window function $W_I$, see eq.~\eqref{eq:WI}, describes a graybody spectrum, where the exponential decline is replaced with a power-law
\begin{eqnarray}
    f_\nu = \left\{ \begin{array}{ll} 
    \left[ \exp\left( \frac{h \nu}{k_B T} \right) - 1 \right]^{-1} \nu^{\beta + 3} & \left( \nu \ll \nu' \right) \\
    \left[ \exp\left( \frac{h \nu'}{k_B T} \right) - 1 \right]^{-1} \nu'^{\beta + 3} \left( \frac{\nu}{\nu'} \right)^{-\alpha} & \left( \nu > \nu' \right)
    \end{array} \right.
\end{eqnarray}
  where $\nu'$ is defined such that the two functions match at $\nu'$ with a smooth gradient.
  In our analysis we fix the parameters of $W_I$ to $z_c= \sigma_z= \beta=\alpha=2$, $T=34 {\rm K}$ and $\nu'= 353 {\rm GHz}$. 
For the CIB noise we consider two components: shot-noise and galactic dust emission,
\begin{equation}
    N^I_L =   N^I_{\rm shot} +   N^I_{\rm dust} (L)
\end{equation}
with $N^I_{\rm shot}=225.6 \times 10^{-12} \ {\rm MJy^2 sr^{-1}}$ and $N^I_{\rm dust} (L)=\left( 3 \times 10^{-4} \right) L^{-2.17}  \ {\rm MJy^2 sr^{-1}}$

\newcommand{\hn}[0]{\ensuremath{\bn}}
\newcommand{\vn}[0]{\ensuremath{\boldsymbol{n}}}
\newcommand{\vs}[0]{\ensuremath{\boldsymbol{s}}}

\subsection{Bispectrum of \texorpdfstring{$\om$}{} with LSS tracers}
\label{app:bisp}
Here we derive the bispectrum between the lensing curl and the LSS probes in some detail. The Fourier transform of the lensing curl $\omega$ is, see eq.~\eqref{eq:omega},
\begin{eqnarray}
    \omega \left( \bL \right) &=&  2 \int \frac{d^2 L_1}{2 \pi } \bL_1 \times \bL \left( \bL_1 \cdot \left( \bL - \bL_1 \right) \right)
    \nonumber \\
    && \times
    \int_0^{r_*}dr\frac{r_*-r}{r_*r} \int_0^{r}dr'\frac{r-r'}{rr'}
    \Phi \left( \bL_1 , t_0-r \right)\Phi \left( \bL- \bL_1 , t_0 -  r' \right) \,.
\end{eqnarray}
We consider LSS tracers of the form
\begin{equation}
    a^j \left( \bL \right) = \int dr W_j \left( r \right) \delta \left( \bL , r \right)
\end{equation}
where $W_j(r)$ is the window function for the tracer.
We compute 
\begin{eqnarray}
    &&  \hspace{-0.5cm}\langle   \omega \left( -\bL \right) a^{i} \left( \bL_1\right) a^j \left(\bL_2 \right) \rangle 
    \nonumber \\ 
&=&
       2 \int \frac{d^2 L'}{2 \pi } \bL' \times \bL \left( \bL' \cdot \left( \bL + \bL' \right) \right)
    \int_0^{r_*}dr\frac{r_*-r}{r_*r} \int_0^{r}dr'\frac{r-r'}{rr'}
    \int dr_1 W_{i} (r_1 )  \int dr_2 W_{j} (r_2 )
    \nonumber \\
    &&
 \langle    \Phi \left( \bL' , t_0-r \right)\Phi \left(- \bL- \bL' , t_0 -  r' \right) \delta\left(\bL_1, r_1 \right) \delta\left(\bL_2, r_2 \right)   \rangle
 \nonumber \\
 &=&
       - 2  \left( - \frac{3}{2} \Omega_M H_0^2 \right)\int \frac{d^2 L'}{2 \pi } \frac{\bL' \times \bL}{\left( L' \right)^2 \left| \bL+ \bL' \right|^2}  \left( \bL' \cdot \left( \bL + \bL' \right) \right)
    \int_0^{r_*}dr\frac{r_*-r}{r_*r} \frac{r^2}{a\left(t_0 - r \right)}
    \nonumber \\
    &&
        \int dr_1 W_{i} (r_1 )  \int dr_2 W_{j} (r_2 )
 \langle    \delta \left( \bL' , r \right)\kappa \left(- \bL- \bL' ,r \right) \delta\left(\bL_1, r_1 \right) \delta\left(\bL_2, r_2 \right)   \rangle
   \nonumber \\
 &=&  \frac{2}{2\pi}  \left( - \frac{3}{2} \Omega_M H_0^2 \right) \delta_D \left(- \bL + \bL_1 + \bL_2 \right)
  \frac{\bL \times \bL_1}{L_1^2 L_2^2}  \left( \bL_1 \cdot \bL_2 \right)
   \nonumber \\ && \times
   \int_0^{r_*}dr\frac{r_*-r}{r_*r} \frac{r^2}{a\left(t_0 - r \right)} C^{\kappa j}_{L_2}(r) C^{\delta i}_{L_1}(r)
   \nonumber \\
   &+&  
    \frac{2}{2\pi}  \left( - \frac{3}{2} \Omega_M H_0^2 \right) \delta_D \left( -\bL + \bL_1 + \bL_2 \right)
  \frac{\bL \times \bL_2}{L_1^2 L_2^2}  \left( \bL_1 \cdot \bL_2 \right)
     \nonumber \\ && \times
   \int_0^{r_*}dr\frac{r_*-r}{r_*r} \frac{r^2}{a\left(t_0 - r \right)} C^{\delta j}_{L_2}(r) C^{\kappa i}_{L_1}(r)
        \nonumber \\
         &=&\frac{2}{2\pi}  \left( - \frac{3}{2} \Omega_M H_0^2 \right) \delta_D \left( -\bL + \bL_1 + \bL_2 \right)
  \frac{\bL_2 \times \bL_1}{L_1^2 L_2^2}  \left( \bL_1 \cdot \bL_2 \right)
  \nonumber \\
  && \times
   \int_0^{r_*}dr\frac{r_*-r}{r_*r} \frac{r^2}{a\left(t_0 - r \right)} \left[ C^{\kappa j}_{L_2}(r) C^{\delta i}_{L_1}(r)
   - 
      C^{\delta j}_{L_2}(r) C^{\kappa i}_{L_1}(r)
  \right]
   \nonumber \\
         &\simeq& - \frac{2}{2\pi}  \left( - \frac{3}{2}\Omega_M H_0^2 \right) \delta_D \left( -\bL + \bL_1 + \bL_2 \right)
  \frac{\bL_1 \times \bL_2}{L_1^2 L_2^2}  \left( \bL_1 \cdot \bL_2 \right)
  \nonumber \\
  && \times
   \int_0^{r_*}dr\frac{r_*-r}{r r_*} \frac{1}{a\left(t_0 - r \right)} \left[ C^{\kappa j}_{L_2}(r) W_{i}(r) P_{\delta \delta} \left( \frac{L_1}{r},r \right)
   - 
     C^{\kappa i}_{L_1}(r)  W_{j}(r) P_{\delta \delta} \left( \frac{L_2}{r},r \right)
  \right]
  \nonumber \\
\end{eqnarray}
where we have used
\begin{eqnarray}
    \kappa \left( \bL, r \right) &=&- L^2 \int_0^r dr' \frac{r -r'}{r r'} \Phi \left( \bL, t_0 - r' \right) \,
\end{eqnarray}
and the Poisson equation within the Limber approximation,
\bea
    \Phi \left( \bL, r \right) &=& - \frac{3 \Omega_M H_0^2}{2 L^2} \frac{r^2}{a} \delta \left( \bL, r \right) \,. 
\eea
We have also introduced the cross spectra of the tracers $i$ with the convergence $\ka$ and the density field $\de$,
\begin{align}
   \int dr' W_{i} (r') \langle \kappa \left( \bL_1 ,r\right) \delta \left( \bL_2 ,r'\right) \rangle  &= \delta_D \left( \bL_1 +\bL_2 \right) C_{L_1}^{\kappa i} \left( r \right) \, \\
   \nonumber \int dr' W_{i} (r') \langle \delta \left( \bL_1 ,r\right) \delta \left( \bL_2 ,r'\right) \rangle  &= \delta_D \left( \bL_1 +\bL_2 \right) C_{L_1}^{\delta i} \left( r \right)
   \nonumber
    \\ & 
   \simeq 
     \delta_D \left( \bL_1 +\bL_2 \right) \frac{W_{i}(r)}{r^2} P_{\delta \delta} \left( \frac{L_1}{r} , r \right) \, .
\end{align}
As in Ref.~\cite{Robertson:2023xkg}, we introduce 
\begin{align}
\label{eq:M_coeff_bisp}
 &   M^{ij} \left( L_1, L_2 \right) =   \frac{3}{2}\Omega_M H_0^2
    \int_0^{r_*}dr\frac{r_*-r}{r r_*} \frac{1}{a\left(t_0 - r \right)} C^{\kappa j}_{L_2}(r) W_{i}(r) P_{\delta \delta} \left( \frac{L_1}{r},r \right)
\end{align}
which yields
\begin{align}
&   b_{(-\bL) \bL_1 \bL_2}^{\omega i j}
          =  2  \frac{\bL_1 \times \bL_2}{L_1^2 L_2^2} 
         \left( \bL_1 \cdot \bL_2 \right)
 \left[ M^{ij} \left( L_1,L_2 \right) -  M^{ji} \left( L_2,L_1 \right)\right] \, .
\end{align}

\section{Details on squeezed limits}\label{a:squeeze}
\newcommand{\Cov}[0]{\text{Cov}}
\newcommand{\Tr}[0]{\text{Tr}}

We justify here some statements made in the main text in Sec.~\ref{ss:squeeze}, in particular on the squeezed (low-$L$) limits, and on the $B$-modes and $EB$ spectrum created by polarization and lensing rotation.

To obtain  squeezed limits of a quadratic estimator in curved-sky geometry it is possible to start with the expressions for the quadratic estimators, compute their reconstruction noise, and expand systematically in $\ell / L$. For example, Ref.~\cite{Carron:2024mki} did proceed in this way and the low-$L$ limits (accurate up to order $(\ell/L)^2$) of all lensing quadratic estimators are listed there (Appendix D).  That appendix also argues that all such limits can be obtained easily from flat-sky calculations of the linear change of the local power spectrum $C(\vl)$ in the presence of a larger mode of the anisotropy source, up to prefactors for spin-weighted fields (like the lensing shear) that are easy to compute. We only  review here for completeness the response of the CMB fields under a locally constant rotation and a locally constant lensing magnification matrix.

For polarization rotation  ($_{\pm 2}P \rightarrow e^{\mp 2i \beta} {}_{\pm 2}P $), and according to our flat-sky conventions, the linear change in $E$ and $B$ mode is given by
\begin{equation}
	\delta (E(\vl) \pm iB(\vl)) = \mp 2 i \beta (E(\vl) \pm iB(\vl)).
\end{equation}
The $B$-modes created by this rotation are $\delta B(\vl) = - 2\beta E(\vl)$. The linear response of the spectra to $\beta$ is given by
\begin{equation}
	\delta C_\ell^{EB} \sim -2 \beta (C_\ell^{EE} - C_\ell^{BB}) ,\quad \delta C_\ell^{TB} \sim -2 \beta C_\ell^{TE},
\end{equation}
and zero for the the other spectra.

With the notation defined in the introduction, the relation between the observed ($\hat n$) and unlensed ($\hat n'$) angles under a locally constant magnification matrix is given by
\begin{align}
	\hat n &= \begin{pmatrix}
		1 + \kappa + \gamma_1 & \gamma_2 - \omega \\ \gamma_2 + \omega & 1 + \kappa - \gamma_1
	\end{pmatrix} \hat n'
 \equiv M \hat n'.
\end{align}
The modes of the polarization fields will be given by
\begin{equation}
	-(E(\vl) \pm iB(\vl)) = \int d^2\hat n \:  {}_{\pm s}P(M^{-1}\hat n) e^{-i \vl \cdot \hat n \mp i s \phi_{\vl}}.
\end{equation}
Expanding, and using $D_{ij} = \delta_{ij} - M_{ij}$, the change to the modes to first order is given by \begin{equation}
	-\delta (E(\vl) \pm iB(\vl)) = \int d^2\hat n \: \left[ D_{ij}\hat n_j\partial_{n_i} {}_{\pm s}P(\hat n)\right] ) e^{-i \vl \cdot \hat n \mp i s \phi_{\vl}}.
\end{equation}
After some simple manipulations, one can write
\begin{align}\label{eq:dEB}
	&\delta (E(\vl) \pm iB(\vl)) =
 \left[ 2\kappa - \vl \cdot D \nabla_{\vl} +  e^{\pm i s \phi_{\vl}}\vl \cdot D \nabla_{\vl}e^{\mp i s \phi_{\vl}} \right] (E(\vl) \pm i B(\vl)), 
	\end{align}
	
\begin{figure}[!ht]
\centering
 	\includegraphics[width=\columnwidth]{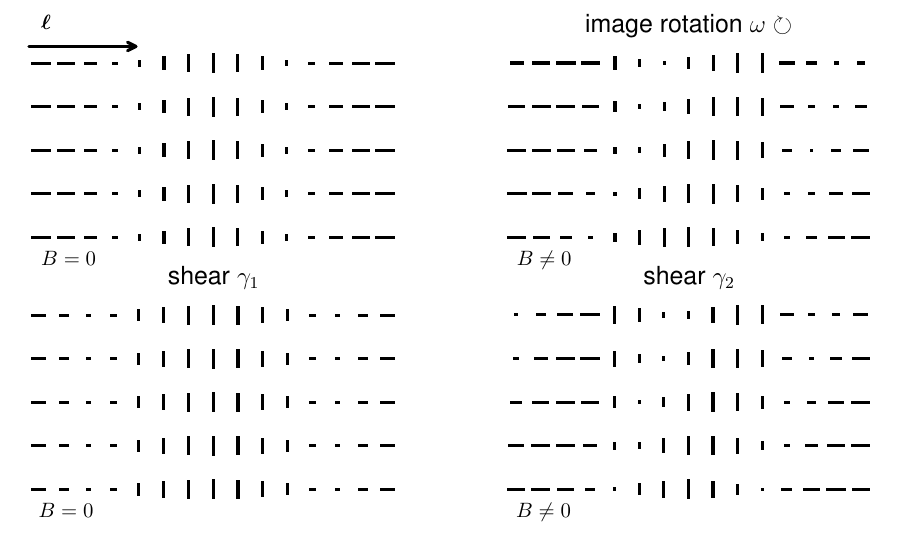}%plotN0.py
 	\caption{\label{f:shear}
$B$-modes from large shearing lenses: The upper left panel shows a pure $E$ polarization pattern, formed by a single $E$-mode $E(\vl)$ with $\vl$ here along the horizontal axis. This results in pure Q, oscillating pattern. The polarization sticks are either horizontal ($Q > 0$) or vertical $(Q < 0)$, with strength $|Q|$ (the sticks length).  A large-scale shearing lens introduces in it a $B$-mode  when their respective angles are sufficiently misaligned (here, 45 degrees for $\gamma_2$, bottom right): in this case the polarization strength also varies at a 45 degrees angle with respect to the sticks. In this maximal case the $B$ pattern created is the same (or opposite) as that of a large-scale image rotation of the same amplitude (top right, for which $\omega = -\gamma_2$). None is produced when their angles are aligned (bottom left, $\gamma_1$)}
\end{figure}

Following our conventions for the magnification matrix, and introducing $ \gamma_1 + i \gamma_2 = \gamma e^{i 2\phi_\gamma}$, we can write
\begin{align}  \label{eq:Del}
	-\vl \cdot D \nabla_{\vl} &= \left[\kappa + \gamma\cos (2(\phi_\gamma - \phi_{\vl}))\right]\: \ell\partial_\ell 
 -  \left[\omega - \gamma\sin (2(\phi_\gamma - \phi_{\vl}))\right]\partial_{\phi_{\vl}}\,.
\end{align}
In particular, this implies that the $B$ modes created from a pure $E$ pattern are given by
\begin{equation}
	\delta B(\vl) = \left[-2\omega + 2\gamma\sin (2(\phi_\gamma - \phi_{\vl}))\right]  E(\vl), \text{  (from lensing rotation and shear)}.
\end{equation}
Clearly, when $\phi_\ga=\phi_{\vl}$ no B-mode is generated from shear (bottom left panel in Fig.~\ref{f:shear}) while for $\phi_\ga-\phi_{\vl}=\pm\pi/4$, the B-mode generation is maximal and equal to that of rotation with $\om =\mp\ga$ (right panels in Fig.~\ref{f:shear}). 
The contribution from lensing rotation has the same sign in these conventions as for polarization rotation.
The changes to the spectra are easily obtained from Eqs~\eqref{eq:dEB} and~\eqref{eq:Del}. The auto-spectra for example obey
\begin{equation}
	\delta C^{XX}_{\vl} = C^{XX}_{\ell} \left[ \kappa \frac{d \ln \ell^2 C^{XX}_\ell}{d \ln \ell} + \gamma\cos (2(\phi_\gamma - \phi_{\vl}))\frac{d \ln  C^{XX}_\ell}{d \ln \ell}   \right]
\end{equation}
for $X \in T, E, B$.
The $EB$ and $TB$ spectra become
\begin{align} 
    \delta C^{EB}_{\vl} &= -(C_\ell^{EE} - C_\ell^{BB})\left(2\omega - 2\gamma \sin(2(\phi_\gamma - \phi_{\vl}))\right)\, , \\
    \delta C^{TB}_{\vl} &= -C_\ell^{TE}\left(2\omega - 2\gamma \sin(2(\phi_\gamma - \phi_{\vl}))\right) \, .
\end{align}
All the squeezed limits given in this paper follow from the recipe of~\cite{Carron:2024mki} appendix D in combination with the  formulae for the linear spectral responses. This is consistent with a similar calculation in~\cite{Prince:2017sms} (after accounting for the different sign convention in $\kappa$ and $\gamma$).

\section{Curved-sky QE responses and biases}\label{app:biases}
\newcommand{\M}[0]{\mathsf{M}}
In this appendix we review briefly the relevant QE responses $(\mathcal R_L)$ and biases ($N^{(0)}$, $N^{(1)}$) in the curved sky. We mostly follow \cite{Ade:2015zua}, with a more general discussion of the curved-sky $N^{(1)}_{L}$. The curved-sky $TT$ lensing $N^{(1)}$ is also discussed in \cite{Hanson:2010rp}, and more generally the structure of statistically isotropic CMB trispectra has been worked out from first principles in \cite{Hu:2001fa}. We also obtain a simple (`squeezed') approximation for the low-$L$ $N^{(1)}$.

Our starting point is the generic linear response of the CMB spectra to a source of anisotropy $\phi_{LM}$, once the relevant symmetries have been extracted
\begin{equation}\label{eq:f}
	\Delta\:\av{X_{l_1m_2}Y_{l_2m_2}} =  \sum_{LM}(-1)^M  \begin{pmatrix}
		\ell_1 & \ell_2 & L \\ m_1 & m_2 & -M
	\end{pmatrix}f^{\phi XY}_{\ell_1\ell_2L}\phi_{LM},
\end{equation}
where $X$ and $Y$ are either $T, E$ or $B$. The isotropic response functions $f^{\phi XY}$ are either purely imaginary for a parity-odd combination of fields (for example $\phi EB$, if $\phi$ is the lensing gradient potential), in which case $f$ is non-zero for $\ell_1 + \ell_2 + L$ an odd number, and real for a parity-even combination, with $\ell_1 + \ell_2 + L$ and even number (for example for $\Omega EB$, if $\Omega$ is the lensing curl potential). The following relations are useful, $f^{XY}_{\ell_1\ell_2L} = (-1)^{L + \ell_1 + \ell_2} f^{YX}_{\ell_2\ell_1L} = (-1)^{L + \ell_1 + \ell_2} f^{XY, *}_{\ell_1\ell_2L}$ and follow directly from \eqref{eq:f}.

A generic quadratic estimator targeting a statistically isotropic anisotropy source may be written as

\begin{equation}\label{eq:qe}
\bar x_{LM}[X, Y] = \frac 12\sum_{\ell_1 m_1 \ell_2 m_2}(-1)^M \begin{pmatrix}
	\ell_1 & \ell_2 & L \\ m_1 & m_2 &-M
\end{pmatrix}	g^{xXY}_{\ell_1\ell_2L} X_{\ell_1m_1}   Y_{\ell_2m_2}.
\end{equation}
Ref.~\cite{Ade:2015zua} defines QE acting on the inverse-variance filtered CMB maps $\bar X$ instead while we define them here acting on the maps $X$ themselves. The former is indeed more natural, but the latter spares us to introduce additional notation in this appendix.

The linear response of $\bar x_{LM}$ to $\phi_{L'M'}$ is obtained by plugging \eqref{eq:f} into \eqref{eq:qe}. The orthogonality of the Wigner 3j symbols forces $L=L'$ and $M=M'$. The result is
\begin{equation}
	\mathcal R^{x \phi}_{L} = \frac{1}{2(2L + 1)} \sum_{\ell_1\ell_2} g^{xXY}_{\ell_1\ell_2L}f^{\phi XY}_{\ell_1\ell_2L}.
\end{equation}
Here the sum goes over the values $\ell_1$ and $\ell_2$ which satisfy the triangle equality, $|\ell_1-\ell_2|\leq L\leq \ell_1+\ell_2$.
The normalized estimator $\bar x_{LM} / \mathcal R_L^{x\phi}$ provides then an estimate of $\phi$.

The covariance of quadratic estimators contains disconnected contractions of the four CMB fields: $\av{XYZW} -\av{XY}\av{ZW} \equiv \av{XZ}\av{YW} + \av{XW}\av{YZ}  + \text{ connected term}$. The disconnected piece can be calculated simply using the orthogonality relations of the Wigner 3j symbols. Let $\bar x[X, Y]$ and $\bar y[Z, W]$ be two QE's. Then their covariance will contain
\begin{align}
	& \hspace{-1cm}\av{\bar x_{LM}\bar y^*_{L'M'}} - \av{\bar x_{LM}}\av{\bar y^*_{L'M'}}
 \nonumber \\
 & \ni\frac {\delta_{LL'}\delta_{MM'}}{4(2L + 1)}\sum_{\ell_1\ell_2}g^{xXY}_{L\ell_1\ell_2}\left[g^{yZW,*}_{L\ell_1\ell_2} C^{XZ}_{\ell_1} C^{YW}_{\ell_2} + (-1)^{\ell_1 + \ell_2 + L}g^{yZW,*}_{L\ell_2\ell_1}C^{XW}_{\ell_1}C^{YZ}_{\ell_2}  \right]  \nonumber \\
	&\equiv \delta_{LL'}\delta_{MM'} n^{(0)}_{L,  x  y}  \text{  ($+$ connected 4-point),}
\end{align}
(the phase in the second term comes from switching the first two columns of the relevant Wigner symbol). The spectra that enter this equation are the total CMB spectra, including noise and the effect of the anisotropy sources. Physically, this represents the reconstruction noise sourced by the total fluctuations of the CMB fields.
The usual $N^{(0)}$ bias is then that of the normalized estimators
\begin{equation}
	N^{(0)\phi}_{L,xy} \equiv \frac{n^{(0)}_{L, x y}}{\mathcal R_L^{x \phi}\mathcal R_L^{y \phi}}.
\end{equation}
We now turn to the connected piece entering such a cross-spectrum. To leading `explicit order' in the anisotropy source spectrum $C_L^{\phi\phi}$, we only need to consider the linear response \eqref{eq:f} (naively, one might consider crossing the second order response on one of the estimators to the zeroth order on the second. This is however a disconnected piece, part of the change in the isotropic spectra from the anisotropy source, already accounted for by $n_L^{(0)}$).
There are 3 ways to produce two factors~\eqref{eq:f} from the four fields formed by a pair of QE's: one way is having the CMB contract on each QE separately ($\Delta \av{XY} \Delta \av{ZW}$, `primary' contractions), and the other two contracting across the two QEs ($\Delta \av{XZ} \Delta \av{YW}$ and $\Delta \av{XW} \Delta \av{YZ}$, `secondary' contractions). The primary term is the most natural term: the normalized $\bar x $ and $\bar y$ were built precisely to form an estimate of $\phi$, hence we ought to find a term proportional to $C_L^{\phi\phi}$. The two other terms are called $n^{(1)}$, or $N^{(1)}$ after normalization of the estimators.

As a warm-up for the $n^{(1)}$ terms, the explicit calculation of the primary term gives indeed the expected result (suppressing summation indices)
\begin{align}
& \hspace{-1cm}\av{\bar x_{LM} \bar y_{L'M}^*}\text{  (primary contractions)} 
\nonumber \\
&= \nonumber\frac 14\sum g^{x XY}_{\ell_1\ell_2L}g^{y ZW*}_{\ell_3\ell_4L'}(-1)^{M-M'}f^{\phi XY}_{\ell_1\ell_2\ell_2} f^{\phi ZW*}_{\ell_3\ell_4\ell_3} (-1)^{M_2 + M_3}\av{\phi_{L_2M_2} \phi^*_{L_3M_3}} \times  \\&\begin{pmatrix}
	\ell_1 & \ell_2 & L \\ m_1 & m_2 & -M
\end{pmatrix}  \begin{pmatrix}
	\ell_3 & \ell_4 & L^{'} \\ m_3 & m_4 & -M^{'}
\end{pmatrix}
\begin{pmatrix}
	\ell_1 & \ell_2 & L_2 \\ m_1 & m_2 & -M_2
\end{pmatrix}   \begin{pmatrix}
	\ell_3 & \ell_4 & L_3 \\ m_3 & m_4 & -M_3
\end{pmatrix} \nonumber
\\ &= \frac{\delta_{LL'}\delta_{MM'}}{4(2L + 1)^2}\sum g^{x XY}_{\ell_1\ell_2L}g^{y ZW*}_{\ell_3\ell_4L}f^{\phi XY}_{\ell_1\ell_2L} f^{\phi ZW*}_{\ell_3\ell_4L} =\delta_{LL'}\delta_{MM'}
	\mathcal R_L^{\phi \bar x}\mathcal R_{L}^{\phi \bar y *} C_L^{\phi\phi} \,.  
\end{align}
We have used the orthogonality of the 3j symbols, when summing over $m_1,m_2$ and $m_3,m_4$.
The secondary terms ($n^{(1)}$) are more complicated. The structure is the  same as on this equation, but the second pair of 3j's originating from the responses will couple now $\ell_1$ and $\ell_2$ to $\ell_3$ and $\ell_4$ and vice-versa. The resulting pattern is precisely that of a 6j Wigner symbol~\cite{Varshalovich, Hu:2001fa, Hanson:2010rp}. After tedious matching of various signs, one obtains the result
\begin{empheq}{align}
	n^{(1)}_{L, xy} &= (-1)^P\frac 1{4(2L + 1)}\sum_{\ell_1\ell_2\ell_3\ell_4 L'}(-1)^{L-L'}g^{xXY}_{\ell_1\ell_2L}g^{yZW}_{\ell_3\ell_4L} C_{L'}^{\phi\phi}
\nonumber \\
& \qquad \times 
 \left[ f^{\phi XW}_{\ell_1\ell_4L'}f^{\phi YZ}_{\ell_2\ell_3L'} \begin{Bmatrix}
		\ell_1 &\ell_2 & L \\ \ell_3 & \ell_4 & L'
	\end{Bmatrix} + f^{\phi XZ}_{\ell_1\ell_3L'}f^{\phi YW}_{\ell_2\ell_4L'} \begin{Bmatrix}
		\ell_1 &\ell_2 & L \\ \ell_4 & \ell_3 & L'
	\end{Bmatrix} \right]
\end{empheq}
where we extracted an overall sign $(-1)^P = (-1)^{\ell_2 + \ell_3 + L'}$, the parity of $\phi YZ$.

Numerical evaluation of the curved-sky formula is tedious, and we have used the corresponding flat-sky expressions for numerical work, available for example in~\cite{Aghanim:2018oex, Maniyar:2021msb}. For spin-0 estimators (like polarization rotation), we can however compute the low-$L$ $n^{(1)}_L$ without too much trouble: for $L=0$ we use~\cite{Varshalovich}
\begin{equation}
	\begin{Bmatrix}
		\ell_1 &\ell_2 & 0 \\ \ell_3 & \ell_4 & L'
	\end{Bmatrix} = (-1)^{\ell_1 + \ell_4 + L'} \frac{\delta_{\ell_1\ell_2}\delta_{\ell_3\ell_4}}{\sqrt{2\ell_1 + 1} \sqrt{2\ell_3 + 1}}.
\end{equation}
After some  rearranging of the terms, we find
\begin{eqnarray}\label{eq:n10}
	n^{(1)}_{L=0, x y} &=& \frac 14\sum_{\ell_1\ell_2 L'}\left(\frac{(-1)^{\ell_1}g^{xXY}_{\ell_1\ell_10}}{\sqrt{2\ell_1 + 1}} \right)\left(\frac{(-1)^{\ell_2}g^{yZW}_{\ell_2\ell_20}}{\sqrt{2\ell_2 + 1}} \right)C_{L'}^{\phi\phi}\times
    \nonumber \\
    && \hspace{2cm} \left[ (-1)^P f^{\phi XW}_{\ell_1\ell_2L'}f^{\phi YZ}_{\ell_1\ell_2L'}  + (-1)^{P'}f^{\phi XZ}_{\ell_1\ell_2L'}f^{\phi YW}_{\ell_1\ell_2L'} \right]
\end{eqnarray}
where the constant signs $(-1)^P$ and $(-1)^{P'}$ are the parity of $\phi YZ$ and $\phi XZ$ respectively. On one hand, this may be evaluated very efficiently in position-space in the exact same manner as usual for response calculations. This is a special case of a more general result valid when both $L$ and $L'$ are small compared to the `small scale mode' $2\ell + 1 \equiv (\ell_1 + \ell_2 + \ell_3 + \ell_4)/2$, which we discuss next.
Wigner 6j symbols have a well-known useful graphical representation through tetrahedra and their value closely follows their inverse volume~\cite{osti_4824659} when momenta are large. From this, one can infer~\cite{Varshalovich} that forcing $L$ and $L'$ to be somewhat small forces in the first term $ 4q\equiv \ell_1 + \ell_3  - (\ell_2 + \ell_4) = 0 $ to dominate. (and  similarly $4q'\equiv \ell_1 + \ell_4 - (\ell_2 + \ell_3) = 0$ in the second term). 
Non-zero values of $q$ are then suppressed by a factor of $(2\ell + 1)^{|4q|}$.
This results eventually in
\begin{equation}\label{eq:n1sq}
n^{(1)}_{LL'}\sim	\frac{1}{4(2L + 1)} C^{\phi\phi}_{L'} \sum_{\ell_1\ell_2\ell_3\ell_4}\frac{ g^{XY}_{\ell_1 \ell_2 L}g^{ZW}_{\ell_3\ell_4 L}}{2\ell +1}\left[\delta_{q0}(-1)^Pf^{XW}_{\ell_1 \ell_4 L'}f^{ZY}_{\ell_3 \ell_2 L'} + \delta_{q'0}(-1)^{P'}f^{XZ}_{\ell_1 \ell_3 L'}f^{WY}_{\ell_4 \ell_2 L'}\right]
\end{equation}
where $n^{(1)}_{LL'}$ is the contribution per $L'$ to $n^{(1)}_L$.
Eqs~\eqref{eq:n10} and~\eqref{eq:n1sq} sums a term which has a similar formal structure as the standard QE response $\mathcal R_{L'}$, but using modified weights. Since all bispectra $f$ and $g$ at small $L$ and $L'$ and large CMB $\ell$'s captures the response of the small scale spectrum to the anisotropy source, one can derive a similarly handy formula in this limit. For example in this way: the polarization-rotation $n^{(1)}_{L\beta\beta}$ induced by the lensing potential $\psi$  will be dominated at low $L$ by the parity even $\beta EB$ (\eqref{eq:n10} explicitly enforces $xXY $ and $yZW$ to be parity-even). Hence $X=Z=E$ and $Y=W=B$, and we can also drop the second term, since the lensing $\psi BB$ linear response is negligible. The relevant set of weights and responses are thus
\begin{align}
	 g_{\ell_1\ell_2 L}^{\beta EB} &=- 2 \sqrt{\frac{(2\ell_1 + 1)(2\ell_2 + 1)(2' + 1)}{4\pi}}\frac{2
     C_{\ell_1}^{EE}}{C_\ell^{EE, \rm dat} C_\ell^{ BB, \rm dat}} {}_{2}F^+_{\ell_1\ell_2 L} \,,\\
	 f^{\psi EB}_{\ell_1\ell_2L'} &= \frac {i} 4 \sqrt{\frac{(2\ell_1 + 1)(2\ell_2 + 1)(2L' + 1)}{4\pi}} (2C_{\ell_1}^{EE}) {}_{2}F^{-}_{\ell_1\ell_2L'} \left(L'(L' + 1) + \ell_1(\ell_1 + 1) - \ell_2(\ell_2 + 1) \right) .
\end{align}
We may then proceed following~\cite{Carron:2024mki} by defining $\M' = \ell_1 - \ell_2$ and $\ell = (\ell_1 + \ell_2)/2$ and work to $\mathcal O( (L/\ell)^2)$. In that order, it helps to notice that $\M'$ is constrained to $|\M'| \leq L'$ from the triangle conditions, that products of smooth functions $v$ like $v_{\ell_1}v_{\ell_2}$ are equivalent to $v_\ell^2$, and that $\ell_2(\ell_1 + 1) - \ell_1(\ell_1 + 1) = \M' (2\ell + 1) $. \cite{Carron:2024mki} gives the squeezed limit of the trigonometric ${}_{2}F$ kernels as function of $\M'$, and the resulting $\M'$-average can be performed using the same methods as in that reference. The end result for the $\psi$-induced $n^{(1)}$ is
\bea
 n^{(1)}_{LL',\beta \beta} &\sim& \left( \frac{2L' + 1}{4\pi} C_{L'}^{\kappa \kappa} \frac 12 \frac{(L'-1)(L' + 2)}{L'(L' +1)}\right) \sum_{2\ell \ge L,L'}  \frac{2\ell + 1}{4\pi}\left(\frac{-2C_\ell^{EE}}{C_\ell^{EE, \rm dat}}\frac{-2C_\ell^{EE}}{C_\ell^{BB, \rm dat}} \right)^2, \nonumber \\
 &&
\label{eq:D12}
\eea
where $C_{L}^{\kappa \kappa} = L^2(L + 1)^2 C_{L}^{\psi\psi}/4$. The left-most factor is the variance of $\kappa$, constrained to $L'\leq 2\ell$ by the triangle conditions on $\ell_1,\ell_2,L'$. The multiplicative factor $(L-1)(L + 2)/L/(L + 1) / 2 \sim 1/2$ in the sum is the `shear' factor, accounting for the fact that the $EB$-response only relates to the shear-part of the lensing field (a large scale (de-)magnifying lens does not change the pure-$E$ nature of a pure-$E$ primordial polarization). This approximation to $n^{(1)}$ does not appear useful in the delensed case, since delensing removes most of the large-scale lensing power, making contributions from large $L'$ most significant.

\bibliographystyle{JHEP}

\begin{thebibliography}{10}

\bibitem{Planck:2018nkj}
{\bf Planck} Collaboration, N.~Aghanim et~al., {\it {Planck 2018 results. I.
  Overview and the cosmological legacy of Planck}},  {\em Astron. Astrophys.}
  {\bf 641} (2020) A1, [\href{http://arxiv.org/abs/1807.06205}{{\tt
  arXiv:1807.06205}}].

\bibitem{Planck:2018lbu}
{\bf Planck} Collaboration, N.~Aghanim et~al., {\it {Planck 2018 results. VIII.
  Gravitational lensing}},  {\em Astron. Astrophys.} {\bf 641} (2020) A8,
  [\href{http://arxiv.org/abs/1807.06210}{{\tt arXiv:1807.06210}}].

\bibitem{Planck:2018vyg}
{\bf Planck} Collaboration, N.~Aghanim et~al., {\it {Planck 2018 results. VI.
  Cosmological parameters}},  {\em Astron. Astrophys.} {\bf 641} (2020) A6,
  [\href{http://arxiv.org/abs/1807.06209}{{\tt arXiv:1807.06209}}]. [Erratum:
  Astron.Astrophys. 652, C4 (2021)].

\bibitem{SPT-3G:2014dbx}
{\bf SPT-3G} Collaboration, B.~A. Benson et~al., {\it {SPT-3G: A
  Next-Generation Cosmic Microwave Background Polarization Experiment on the
  South Pole Telescope}},  {\em Proc. SPIE Int. Soc. Opt. Eng.} {\bf 9153}
  (2014) 91531P, [\href{http://arxiv.org/abs/1407.2973}{{\tt
  arXiv:1407.2973}}].

\bibitem{SPT-3G:2024qkd}
{\bf SPT-3G} Collaboration, K.~Prabhu et~al., {\it {Testing the
  \ensuremath{\Lambda}CDM Cosmological Model with Forthcoming Measurements of
  the Cosmic Microwave Background with SPT-3G}},  {\em Astrophys. J.} {\bf 973}
  (2024), no.~1 4, [\href{http://arxiv.org/abs/2403.17925}{{\tt
  arXiv:2403.17925}}].

\bibitem{ACT:2020gnv}
{\bf ACT} Collaboration, S.~Aiola et~al., {\it {The Atacama Cosmology
  Telescope: DR4 Maps and Cosmological Parameters}},  {\em JCAP} {\bf 12}
  (2020) 047, [\href{http://arxiv.org/abs/2007.07288}{{\tt arXiv:2007.07288}}].

\bibitem{ACT:2023kun}
{\bf ACT} Collaboration, M.~S. Madhavacheril et~al., {\it {The Atacama
  Cosmology Telescope: DR6 Gravitational Lensing Map and Cosmological
  Parameters}},  {\em Astrophys. J.} {\bf 962} (2024), no.~2 113,
  [\href{http://arxiv.org/abs/2304.05203}{{\tt arXiv:2304.05203}}].

\bibitem{SimonsObservatory:2018koc}
{\bf Simons Observatory} Collaboration, P.~Ade et~al., {\it {The Simons
  Observatory: Science goals and forecasts}},  {\em JCAP} {\bf 02} (2019) 056,
  [\href{http://arxiv.org/abs/1808.07445}{{\tt arXiv:1808.07445}}].

\bibitem{LiteBIRD:2022cnt}
{\bf LiteBIRD} Collaboration, E.~Allys et~al., {\it {Probing Cosmic Inflation
  with the LiteBIRD Cosmic Microwave Background Polarization Survey}},  {\em
  PTEP} {\bf 2023} (2023), no.~4 042F01,
  [\href{http://arxiv.org/abs/2202.02773}{{\tt arXiv:2202.02773}}].

\bibitem{LiteBIRD:2023iiy}
{\bf LiteBIRD} Collaboration, A.~I. Lonappan et~al., {\it {LiteBIRD science
  goals and forecasts: a full-sky measurement of gravitational lensing of the
  CMB}},  {\em JCAP} {\bf 06} (2024) 009,
  [\href{http://arxiv.org/abs/2312.05184}{{\tt arXiv:2312.05184}}].

\bibitem{Abazajian:2019eic}
K.~Abazajian et~al., {\it {CMB-S4 Science Case, Reference Design, and Project
  Plan}},  \href{http://arxiv.org/abs/1907.04473}{{\tt arXiv:1907.04473}}.

\bibitem{2022ApJ...926...54A}
{\bf CMB-S4} Collaboration, K.~Abazajian et~al., {\it {CMB-S4: Forecasting
  Constraints on Primordial Gravitational Waves}},  {\em Astrophys. J.} {\bf
  926} (2022), no.~1 54, [\href{http://arxiv.org/abs/2008.12619}{{\tt
  arXiv:2008.12619}}].

\bibitem{Pratten:2016dsm}
G.~Pratten and A.~Lewis, {\it {Impact of post-Born lensing on the CMB}},  {\em
  JCAP} {\bf 1608} (2016), no.~08 047,
  [\href{http://arxiv.org/abs/1605.05662}{{\tt arXiv:1605.05662}}].

\bibitem{Marozzi:2016uob}
G.~Marozzi, G.~Fanizza, E.~Di~Dio, and R.~Durrer, {\it {CMB-lensing beyond the
  Born approximation}},  {\em JCAP} {\bf 09} (2016) 028,
  [\href{http://arxiv.org/abs/1605.08761}{{\tt arXiv:1605.08761}}].

\bibitem{Lewis:2016tuj}
A.~Lewis and G.~Pratten, {\it {Effect of lensing non-Gaussianity on the CMB
  power spectra}},  {\em JCAP} {\bf 12} (2016) 003,
  [\href{http://arxiv.org/abs/1608.01263}{{\tt arXiv:1608.01263}}].

\bibitem{Lewis:2017ans}
A.~Lewis, A.~Hall, and A.~Challinor, {\it {Emission-angle and
  polarization-rotation effects in the lensed CMB}},  {\em JCAP} {\bf 08}
  (2017) 023, [\href{http://arxiv.org/abs/1706.02673}{{\tt arXiv:1706.02673}}].

\bibitem{Namikawa:2021obu}
T.~Namikawa, A.~Naruko, R.~Saito, A.~Taruya, and D.~Yamauchi, {\it {Unified
  approach to secondary effects on the CMB B-mode polarization}},  {\em JCAP}
  {\bf 10} (2021) 029, [\href{http://arxiv.org/abs/2103.10639}{{\tt
  arXiv:2103.10639}}].

\bibitem{Marozzi:2016und}
G.~Marozzi, G.~Fanizza, E.~Di~Dio, and R.~Durrer, {\it {Impact of
  Next-to-Leading Order Contributions to Cosmic Microwave Background Lensing}},
   {\em Phys. Rev. Lett.} {\bf 118} (2017), no.~21 211301,
  [\href{http://arxiv.org/abs/1612.07650}{{\tt arXiv:1612.07650}}].

\bibitem{Marozzi:2016qxl}
G.~Marozzi, G.~Fanizza, E.~Di~Dio, and R.~Durrer, {\it {CMB-lensing beyond the
  leading order: temperature and polarization anisotropies}},  {\em Phys. Rev.
  D} {\bf 98} (2018), no.~2 023535,
  [\href{http://arxiv.org/abs/1612.07263}{{\tt arXiv:1612.07263}}].

\bibitem{DiDio:2019rfy}
E.~Di~Dio, R.~Durrer, G.~Fanizza, and G.~Marozzi, {\it {Rotation of the CMB
  polarization by foreground lensing}},  {\em Phys. Rev. D} {\bf 100} (2019),
  no.~4 043508, [\href{http://arxiv.org/abs/1905.12573}{{\tt
  arXiv:1905.12573}}].

\bibitem{Robertson:2023xkg}
M.~Robertson and A.~Lewis, {\it {How to detect lensing rotation}},  {\em JCAP}
  {\bf 08} (2023) 048, [\href{http://arxiv.org/abs/2303.13313}{{\tt
  arXiv:2303.13313}}].

\bibitem{Durrer:2020fza}
R.~Durrer, {\em {The Cosmic Microwave Background}}.
\newblock Cambridge University Press, 12, 2020.

\bibitem{Hirata:2003ka}
C.~M. Hirata and U.~Seljak, {\it {Reconstruction of lensing from the cosmic
  microwave background polarization}},  {\em Phys. Rev.} {\bf D68} (2003)
  083002, [\href{http://arxiv.org/abs/astro-ph/0306354}{{\tt
  astro-ph/0306354}}].

\bibitem{Lewis:2006fu}
A.~Lewis and A.~Challinor, {\it {Weak gravitational lensing of the cmb}},  {\em
  Phys. Rept.} {\bf 429} (2006) 1--65,
  [\href{http://arxiv.org/abs/astro-ph/0601594}{{\tt astro-ph/0601594}}].

\bibitem{Gorski:2004by}
K.~M. G\'orski, E.~Hivon, A.~J. Banday, B.~D. Wandelt, F.~K. Hansen,
  M.~Reinecke, and M.~Bartelman, {\it {HEALPix - A Framework for high
  resolution discretization, and fast analysis of data distributed on the
  sphere}},  {\em Astrophys. J.} {\bf 622} (2005) 759--771,
  [\href{http://arxiv.org/abs/astro-ph/0409513}{{\tt astro-ph/0409513}}].

\bibitem{Zonca2019}
A.~Zonca, L.~Singer, D.~Lenz, M.~Reinecke, C.~Rosset, E.~Hivon, and K.~Gorski,
  {\it healpy: equal area pixelization and spherical harmonics transforms for
  data on the sphere in python},  {\em Journal of Open Source Software} {\bf 4}
  (Mar., 2019) 1298.

\bibitem{Fanizza_2013}
G.~Fanizza, M.~Gasperini, G.~Marozzi, and G.~Veneziano, {\it An exact jacobi
  map in the geodesic light-cone gauge},  {\em Journal of Cosmology and
  Astroparticle Physics} {\bf 2013} (Nov., 2013) 019--019.

\bibitem{Hu:2001kj}
W.~Hu and T.~Okamoto, {\it {Mass reconstruction with cmb polarization}},  {\em
  Astrophys. J.} {\bf 574} (2002) 566--574,
  [\href{http://arxiv.org/abs/astro-ph/0111606}{{\tt astro-ph/0111606}}].

\bibitem{Okamoto:2003zw}
T.~Okamoto and W.~Hu, {\it {CMB lensing reconstruction on the full sky}},  {\em
  Phys. Rev.} {\bf D67} (2003) 083002,
  [\href{http://arxiv.org/abs/astro-ph/0301031}{{\tt astro-ph/0301031}}].

\bibitem{Maniyar:2021msb}
A.~S. Maniyar, Y.~Ali-Ha\"\i{}moud, J.~Carron, A.~Lewis, and M.~S.
  Madhavacheril, {\it {Quadratic estimators for CMB weak lensing}},  {\em Phys.
  Rev. D} {\bf 103} (2021), no.~8 083524,
  [\href{http://arxiv.org/abs/2101.12193}{{\tt arXiv:2101.12193}}].

\bibitem{Yadav:2009eb}
A.~P.~S. Yadav, R.~Biswas, M.~Su, and M.~Zaldarriaga, {\it {Constraining a
  spatially dependent rotation of the Cosmic Microwave Background
  Polarization}},  {\em Phys. Rev. D} {\bf 79} (2009) 123009,
  [\href{http://arxiv.org/abs/0902.4466}{{\tt arXiv:0902.4466}}].

\bibitem{Gluscevic:2009mm}
V.~Gluscevic, M.~Kamionkowski, and A.~Cooray, {\it {De-Rotation of the Cosmic
  Microwave Background Polarization: Full-Sky Formalism}},  {\em Phys. Rev. D}
  {\bf 80} (2009) 023510, [\href{http://arxiv.org/abs/0905.1687}{{\tt
  arXiv:0905.1687}}].

\bibitem{Carron:2013nfa}
J.~Carron and I.~Szapudi, {\it {Sufficient observables for large scale
  structure in galaxy surveys}},  {\em Mon. Not. Roy. Astron. Soc.} {\bf 439}
  (2014) L11, [\href{http://arxiv.org/abs/1310.6038}{{\tt arXiv:1310.6038}}].

\bibitem{Alsing:2017var}
J.~Alsing and B.~Wandelt, {\it {Generalized massive optimal data compression}},
   {\em Mon. Not. Roy. Astron. Soc.} {\bf 476} (2018), no.~1 L60--L64,
  [\href{http://arxiv.org/abs/1712.00012}{{\tt arXiv:1712.00012}}].

\bibitem{Carron:2024mki}
J.~Carron and A.~Lewis, {\it {Spherical bispectrum expansion and quadratic
  estimators}},  {\em JCAP} {\bf 07} (2024) 067,
  [\href{http://arxiv.org/abs/2404.16797}{{\tt arXiv:2404.16797}}].

\bibitem{Aghanim:2018oex}
{\bf Planck} Collaboration, N.~Aghanim et~al., {\it {Planck 2018 results. VIII.
  Gravitational lensing}},  {\em Astron. Astrophys.} {\bf 641} (2020) A8,
  [\href{http://arxiv.org/abs/1807.06210}{{\tt arXiv:1807.06210}}].

\bibitem{Namikawa:2011cs}
T.~Namikawa, D.~Yamauchi, and A.~Taruya, {\it {Full-sky lensing reconstruction
  of gradient and curl modes from CMB maps}},  {\em JCAP} {\bf 01} (2012) 007,
  [\href{http://arxiv.org/abs/1110.1718}{{\tt arXiv:1110.1718}}].

\bibitem{Smith:2010gu}
K.~M. Smith, D.~Hanson, M.~LoVerde, C.~M. Hirata, and O.~Zahn, {\it {Delensing
  CMB Polarization with External Datasets}},  {\em JCAP} {\bf 06} (2012) 014,
  [\href{http://arxiv.org/abs/1010.0048}{{\tt arXiv:1010.0048}}].

\bibitem{CMB-S4:2020lpa}
{\bf CMB-S4} Collaboration, K.~Abazajian et~al., {\it {CMB-S4: Forecasting
  Constraints on Primordial Gravitational Waves}},  {\em Astrophys. J.} {\bf
  926} (2022), no.~1 54, [\href{http://arxiv.org/abs/2008.12619}{{\tt
  arXiv:2008.12619}}].

\bibitem{Belkner:2023duz}
{\bf CMB-S4} Collaboration, S.~Belkner, J.~Carron, L.~Legrand, C.~Umilt\`a,
  C.~Pryke, and C.~Bischoff, {\it {CMB-S4: Iterative Internal Delensing and r
  Constraints}},  {\em Astrophys. J.} {\bf 964} (2024), no.~2 148,
  [\href{http://arxiv.org/abs/2310.06729}{{\tt arXiv:2310.06729}}].

\bibitem{Kesden:2003cc}
M.~H. Kesden, A.~Cooray, and M.~Kamionkowski, {\it {Lensing reconstruction with
  CMB temperature and polarization}},  {\em Phys. Rev. D} {\bf 67} (2003)
  123507, [\href{http://arxiv.org/abs/astro-ph/0302536}{{\tt
  astro-ph/0302536}}].

\bibitem{Hu:2001fa}
W.~Hu, {\it {Angular trispectrum of the CMB}},  {\em Phys. Rev. D} {\bf 64}
  (2001) 083005, [\href{http://arxiv.org/abs/astro-ph/0105117}{{\tt
  astro-ph/0105117}}].

\bibitem{Hanson:2010rp}
D.~Hanson, A.~Challinor, G.~Efstathiou, and P.~Bielewicz, {\it {CMB temperature
  lensing power reconstruction}},  {\em Phys. Rev. D} {\bf 83} (2011) 043005,
  [\href{http://arxiv.org/abs/1008.4403}{{\tt arXiv:1008.4403}}].

\bibitem{Cai:2022zad}
H.~Cai, Y.~Guan, T.~Namikawa, and A.~Kosowsky, {\it {Impact of anisotropic
  birefringence on measuring cosmic microwave background lensing}},  {\em Phys.
  Rev. D} {\bf 107} (2023), no.~4 043513,
  [\href{http://arxiv.org/abs/2209.08749}{{\tt arXiv:2209.08749}}].

\bibitem{Cai:2024zau}
H.~Cai, Y.~Guan, T.~Namikawa, and A.~Kosowsky, {\it {Efficient estimation of
  rotation-induced bias to reconstructed CMB lensing power spectrum}},  {\em
  Phys. Rev. D} {\bf 110} (2024), no.~10 103507,
  [\href{http://arxiv.org/abs/2408.13612}{{\tt arXiv:2408.13612}}].

\bibitem{BICEP2:2018kqh}
{\bf BICEP2, Keck Array} Collaboration, P.~A.~R. Ade et~al., {\it {BICEP2 /
  Keck Array X: Constraints on Primordial Gravitational Waves using Planck,
  WMAP, and New BICEP2/Keck Observations through the 2015 Season}},  {\em Phys.
  Rev. Lett.} {\bf 121} (2018) 221301,
  [\href{http://arxiv.org/abs/1810.05216}{{\tt arXiv:1810.05216}}].

\bibitem{Challinor:2005jy}
A.~Challinor and A.~Lewis, {\it {Lensed CMB power spectra from all-sky
  correlation functions}},  {\em Phys. Rev. D} {\bf 71} (2005) 103010,
  [\href{http://arxiv.org/abs/astro-ph/0502425}{{\tt astro-ph/0502425}}].

\bibitem{Fabbian:2017wfp}
G.~{Fabbian}, M.~{Calabrese}, and C.~{Carbone}, {\it {CMB weak-lensing beyond
  the Born approximation: a numerical approach}},  {\em JCAP} {\bf 2018} (Feb.,
  2018) 050, [\href{http://arxiv.org/abs/1702.03317}{{\tt arXiv:1702.03317}}].

\bibitem{Robertson:2024qjl}
M.~Robertson, G.~Fabbian, J.~Carron, and A.~Lewis, {\it {Detectable signals of
  post-Born lensing curl B-modes}},
  \href{http://arxiv.org/abs/2406.19998}{{\tt arXiv:2406.19998}}.

\bibitem{Kosowsky:2004zh}
A.~Kosowsky, T.~Kahniashvili, G.~Lavrelashvili, and B.~Ratra, {\it {Faraday
  rotation of the Cosmic Microwave Background polarization by a stochastic
  magnetic field}},  {\em Phys. Rev. D} {\bf 71} (2005) 043006,
  [\href{http://arxiv.org/abs/astro-ph/0409767}{{\tt astro-ph/0409767}}].

\bibitem{Pogosian:2019jbt}
L.~Pogosian, M.~Shimon, M.~Mewes, and B.~Keating, {\it {Future CMB constraints
  on cosmic birefringence and implications for fundamental physics}},  {\em
  Phys. Rev. D} {\bf 100} (2019), no.~2 023507,
  [\href{http://arxiv.org/abs/1904.07855}{{\tt arXiv:1904.07855}}].

\bibitem{Zhong:2024tgw}
Y.~Zhong, H.~Cai, S.-Y. Li, Y.~Liu, M.~Li, and W.~Fang, {\it {Forecasts on
  Anisotropic Cosmic Birefringence Constraints for CMB Experiment in the
  Northern Hemisphere}},  \href{http://arxiv.org/abs/2409.01098}{{\tt
  arXiv:2409.01098}}.

\bibitem{Eskilt:2022cff}
J.~R. Eskilt and E.~Komatsu, {\it {Improved constraints on cosmic birefringence
  from the WMAP and Planck cosmic microwave background polarization data}},
  {\em Phys. Rev. D} {\bf 106} (2022), no.~6 063503,
  [\href{http://arxiv.org/abs/2205.13962}{{\tt arXiv:2205.13962}}].

\bibitem{Alvarez:2019rhd}
M.~Alvarez et~al., {\it {PICO: Probe of Inflation and Cosmic Origins}},  {\em
  Bull. Am. Astron. Soc.} {\bf 51} (2019), no.~7 194,
  [\href{http://arxiv.org/abs/1908.07495}{{\tt arXiv:1908.07495}}].

\bibitem{Takahashi:2012em}
R.~Takahashi, M.~Sato, T.~Nishimichi, A.~Taruya, and M.~Oguri, {\it {Revising
  the Halofit Model for the Nonlinear Matter Power Spectrum}},  {\em Astrophys.
  J.} {\bf 761} (2012) 152, [\href{http://arxiv.org/abs/1208.2701}{{\tt
  arXiv:1208.2701}}].

\bibitem{LSSTScience:2009jmu}
{\bf LSST Science, LSST Project} Collaboration, P.~A. Abell et~al., {\it {LSST
  Science Book, Version 2.0}},  \href{http://arxiv.org/abs/0912.0201}{{\tt
  arXiv:0912.0201}}.

\bibitem{Hall:2009rv}
N.~R. Hall et~al., {\it {Angular Power Spectra of the Millimeter Wavelength
  Background Light from Dusty Star-forming Galaxies with the South Pole
  Telescope}},  {\em Astrophys. J.} {\bf 718} (2010) 632--646,
  [\href{http://arxiv.org/abs/0912.4315}{{\tt arXiv:0912.4315}}].

\bibitem{Prince:2017sms}
H.~Prince, K.~Moodley, J.~Ridl, and M.~Bucher, {\it {Real space lensing
  reconstruction using cosmic microwave background polarization}},  {\em JCAP}
  {\bf 01} (2018) 034, [\href{http://arxiv.org/abs/1709.02227}{{\tt
  arXiv:1709.02227}}].

\bibitem{Ade:2015zua}
{\bf Planck} Collaboration, P.~A.~R. Ade et~al., {\it {Planck 2015 results. XV.
  Gravitational lensing}},  {\em Astron. Astrophys.} {\bf 594} (2016) A15,
  [\href{http://arxiv.org/abs/1502.01591}{{\tt arXiv:1502.01591}}].

\bibitem{Varshalovich}
D.~A. {Varshalovich}, A.~N. {Moskalev}, and V.~K. {Khersonskii}, {\em {Quantum
  Theory of Angular Momentum}}.
\newblock {World Scientific}, 1988.

\bibitem{osti_4824659}
G.~Ponzano and T.~Regge, {\it Semiclassical limit of racah coefficients.},  in
  {\em Spectroscopic and Group Theoretical Methods in Physics} (F.~Block, ed.),
  pp.~1--58, John Wiley and Sons, 10, 1969.

\end{thebibliography}
\providecommand{\href}[2]{#2}\begingroup\raggedright\endgroup

\end{document}